\documentclass[prd,eqsecnum,superscriptaddress, showpacs,notitlepage, nofootinbib
]{revtex4-1}

\global\arraycolsep=2pt
\usepackage{amsmath}
\usepackage{amssymb}
\usepackage{graphicx}
\usepackage[utf8]{inputenc}
\usepackage{bm}
\usepackage{pstricks}
\usepackage{verbatim}
\usepackage[colorlinks=true,urlcolor=blue,linkcolor=red,citecolor=red]{hyperref}
\usepackage{color}
\usepackage{ulem} 

\DeclareMathOperator{\Tr}{Tr}
\DeclareMathOperator{\tr}{tr}

\begin{document}

\title{Electromagnetic $\delta$-function sphere}
\author{Prachi Parashar} 
\email{prachi.parashar@ntnu.no}
\homepage{https://www.ntnu.edu/employees/prachi.parashar}
\affiliation{Department of Energy and Process Engineering, Norwegian University of Science 
and Technology, NO-7491 Trondheim, Norway}
\affiliation{Homer L. Dodge Department of Physics and Astronomy, University
of Oklahoma, Norman, OK 73019}

\author{Kimball A. Milton}
\email{kmilton@ou.edu}
\homepage{http://www.nhn.ou.edu/~milton}
\affiliation{Homer L. Dodge Department of Physics and Astronomy, University
of Oklahoma, Norman, OK 73019}

\author{K. V. Shajesh}
\email{kvshajesh@gmail.com}
 \homepage{http://www.physics.siu.edu/~shajesh}
\affiliation{Department of Physics, Southern Illinois University-Carbondale,
Carbondale, IL 62901}
\affiliation{Department of Energy and Process Engineering, Norwegian University of Science 
and Technology, NO-7491 Trondheim, Norway}

\author{Iver Brevik}
\email{iver.h.brevik@ntnu.no}
\homepage{http://folk.ntnu.no/iverhb}
\affiliation{Department of Energy and Process Engineering, Norwegian University of Science 
and Technology, NO-7491 Trondheim, Norway}

\begin{abstract}
We develop a formalism to extend our previous work on the electromagnetic 
$\delta$-function plates to a spherical surface.  The electric ($\lambda_e$) and magnetic 
($\lambda_g$) couplings to the surface are through $\delta$-function potentials defining 
the dielectric permittivity and the diamagnetic permeability, with two anisotropic coupling 
tensors. The formalism incorporates dispersion. The 
electromagnetic Green's dyadic breaks up into transverse electric and transverse magnetic 
parts. We derive the Casimir 
interaction energy between two concentric $\delta$-function spheres in this formalism and 
show that it has the correct asymptotic flat plate limit. We systematically 
derive expressions for the Casimir self-energy and the total stress on a 
spherical shell using a $\delta$-function potential, properly regulated by temporal and 
spatial point-splitting, which are different from the conventional temporal point-splitting. 
In strong coupling, we recover the usual result for the perfectly conducting spherical shell 
but in addition there is an integrated curvature-squared divergent contribution. For finite coupling, there are additional divergent contributions; in 
particular, there is a familiar logarithmic divergence occurring in the third order of the 
uniform asymptotic expansion that renders it impossible to extract a unique finite energy 
except in the case of an isorefractive sphere, which translates into 
$\lambda_g=-\lambda_e$.
\end{abstract}

\maketitle

\section{Introduction}
Having established the surprising result that a pair of parallel neutral perfect 
conductors experiences an attractive force due to fluctuations in the quantum 
electromagnetic field~\cite{Casimir:1948dh}, Casimir suggested that this attraction should 
persist for a spherical shell, and could contribute to the stabilization of the electron 
\cite{Casimir:1953qed}.  On the contrary, when Boyer first did the calculation, he found a 
repulsive result~\cite{Boyer:1968uf}, which was confirmed subsequently by many authors, 
for example in 
Refs.~\cite{Davies:1972css,Balian:1977qr,Milton:1978sf,Leseduarte:1996sac,Nesterenko:1998ces}, 
\begin{equation} 
E_{\rm pcs} =\frac{0.04618}a,
\end{equation}
where $a$ is the radius of the perfectly conducting sphere (pcs). This is a rather unique 
result in the litany of Casimir self-energies, in that it is finite and unambiguous, 
resulting from precise cancellations between interior and exterior contributions and 
between transverse electric (TE) and transverse magnetic (TM) modes.  For example, 
although a finite scalar Casimir self-energy for an infinitesimally thin spherical shell 
imposing Dirichlet boundary conditions may be unambiguously extracted~\cite{Leseduarte:1996sac,Bender:1994zr,Milton:1997vsp}, divergent terms are omitted in doing so. And for 
other shapes, such as rectangular~\cite{Lukosz:1971prc,Lukosz:1973ccs,Lukosz:1973ics,Ambjorn:1981xw} or tetrahedral 
\cite{Abalo:2012jz} cavities, only the interior contributions can be included, although a 
unique self-energy can be extracted, exhibiting a universal behavior. In these cases, 
well-known divergences, identified through heat-kernel analyses, remain. The situation 
becomes even murkier with real materials. For example, a dielectric sphere exhibits an 
unremovable logarithmic divergence~\cite{Milton:1979yx, Bordag:1998vs}, which cannot be 
removed even after accounting for dispersion~\cite{Brevik:1987zi}; only when the speed of 
light is the same inside and outside the sphere is the Casimir self-energy 
finite~\cite{Brevik:1982bpp,Brevik:1982iso}. In the 
dilute limit, $\varepsilon-1\ll 1$, where $\varepsilon$ is the permittivity, a finite 
result in the second order of the coupling is extractable 
\cite{Milton:1997ky,Brevik:1998zs}
\begin{equation}
E_{\rm dds}=\frac{23}{1536\pi}\frac{(\varepsilon-1)^2}a.
\end{equation}
In the next order, however, the above-mentioned divergence appears.

Clearly, there are issues still to be understood involving quantum vacuum self-energies. 
In an effort to establish better control over the calculations and at the same time have a 
flexible formulation, we considered diaphanous materials modeled by 
$\delta$-function contributions to the electric permittivity and the magnetic 
permeability in Refs.~\cite{Parashar:2012it,Milton:2013bm}. We considered an 
infinitesimally thin translucent 
plane surface and learned that the permittivity and permeability potentials were 
necessarily anisotropic. Here, we adapt that formalism to spherical geometry; in addition, 
we regulate the frequency integrals and angular momentum sums by introducing temporal and 
spatial point-splitting regulators, which turned out to be extremely effective in 
geometries with curvature and corner divergences~\cite{Milton:2013yqa, Milton:2013xia}. 
Specifically, we keep both temporal and spatial point-splitting cutoffs, which were 
proposed as a tool for a systematic analysis in the context of the principle of 
virtual work in Refs.~\cite{Estrada:2012yn,Fulling:2012wa}.

In this paper, we will work in natural units $\hbar=c=1$. In the next section, we derive 
general formulas for the energy (and free energy at nonzero temperature) when dispersion 
is present. In Sec.~\ref{d-potential}, we summarize the concept of the $\delta$-function 
potential as introduced in Ref.~\cite{Parashar:2012it}. We obtain the non-trivial boundary 
conditions imposed by the $\delta$-function potentials on the fields, in the presence of a 
spherical boundary, from Maxwell's equations in Sec.~\ref{delta-sphere}, and set up 
Green's dyadics for Maxwell's equations, with the appropriate boundary (matching) 
conditions. In Sec.~\ref{con-sph}, we verify the Green's dyadic structure by evaluating 
the Casimir interaction energy between two concentric $\delta$-function spheres, where the 
asymptotic flat plate limit, i.e. the large radius and small angle, reproduces the 
interaction energy between two parallel $\delta$-function plates. (This coincides with 
the proximity force approximation, PFA, for the spherical surfaces.)
For the case of a purely electric potential, contributing only 
to the permittivity, and with the choice of a plasma model to represent the frequency 
dependence of that coupling, we analyze the resulting electromagnetic vacuum energy in 
Sec.~\ref{self-E-el}. We first analyze the self-energy of a $\delta$-function plate for 
both strong and finite coupling. In the strong coupling case, the divergences cancel between 
transverse electric and transverse magnetic mode. However, for the finite coupling case, we 
see a logarithmic divergence appearing in the third order of the couping parameter in 
addition to an inverse power of the point-splitting parameter. For the spherical shell, 
in the strong coupling we recover the familiar result of Boyer~\cite{Boyer:1968uf}, but 
with a divergent term, due to the square of the curvature of the sphere, whose form 
depends on the precise nature of the point-splitting cutoff. This 
divergence is not observed in the conventional temporal point-splitting cutoff. For finite 
coupling, the divergence structure is more complicated, and there emerges the familiar 
logarithmic 
dependence on the cutoff, which first appears in third order in the strength of the 
potential.  Because the scale of this logarithm is ambiguous, no unique finite part can be 
computed.  We verify these results by computing in Sec.~\ref{sec4}, directly from the 
stress tensor, the pressure on the spherical shell. Finally, in Sec.~\ref{sec5}, we see 
how the results are modified when both potentials, electric and magnetic, are included. 
Apart from strong coupling, the only possible finite case is that for isorefractivity, 
when the electric and magnetic coupling are equal in magnitude but opposite in sign,  
corresponding to $\varepsilon\mu=1$.  In the Conclusion, we discuss our results in light 
of recent literature, which might bear on some of the issues raised here. 

We have extended our study to the finite temperature analysis of a $\delta$-function 
shell~\cite{Milton:2017sp-ent}, which shows a discomforting negative entropy behavior in 
addition to the temperature dependent divergences. In the $T=0$ study we can avoid these 
subtleties and gain more insight into the divergence structure depending only on the 
point-splitting cutoff parameter.

\section{Formalism}
\label{formal}
It is convenient to consider the general finite temperature case first. The free energy, 
including the bulk contributions, is 
\begin{equation}
F=-\frac{T}2\sum_{n=-\infty}^\infty \Tr\ln\bm{\Gamma},
\end{equation}
where Green's dyadic for a arbitrary electromagnetic system at temperature $T$ in the 
presence of a dispersive dielectric and diamagnetic  material satisfies the differential 
equation
\begin{equation}
\bm{\Gamma}^{-1}\bm{\Gamma}=\bm{1},\quad
\bm{\Gamma}^{-1}=-\frac1{\zeta_n^2}\bm{\nabla}\times\frac1{\mu(\zeta_n)}
\bm{\nabla}\times-\varepsilon(\zeta_n),\label{degf}
\end{equation}
in terms of the Matsubara frequency $\zeta_n=2\pi n T$. The entropy is 
\begin{equation}
S=-\frac{\partial F}{\partial T}=-\frac{F}T+\frac{U}T,
\end{equation}
from which we identify the internal energy 
\begin{eqnarray}
U&=&\frac{T}2\sum_{n=-\infty}^\infty \zeta_n
\frac\partial{\partial \zeta_n}\Tr\ln
\bm{\Gamma}\nonumber \\
&=&\frac{T}2\sum_{n=-\infty}^\infty \zeta_n\Tr\bm{\Gamma}^{-1}
\frac{\partial}{\partial\zeta_n}\bm{\bm{\Gamma}}.
\end{eqnarray}
The differential equation (\ref{degf}) allows us to transfer the derivative to the first 
factor in the trace, and then subsequently that equation implies 
\begin{eqnarray}
U&=&-\frac{T}2\sum_{n=-\infty}^\infty \zeta_n\Tr\bm{\Gamma}\frac\partial
{\partial\zeta_n}\bm{\Gamma}^{-1}\nonumber\\
&=&-\frac{T}2\sum_{n=-\infty}^\infty
\zeta_n\Tr\bm{\Gamma}\left[\frac2{\zeta_n^3}\bm{\nabla}\times
\frac1{\mu(\zeta_n)} 
\bm{\nabla}\times-\frac\partial{\partial\zeta_n}\varepsilon(\zeta_n)
+\frac1{\zeta_n^2}\bm{\nabla}\times\frac1{\mu^2(\zeta_n)}\frac{\partial
\mu(\zeta_n)}{\partial\zeta_n}\bm{\nabla}\times\right]\nonumber\\
&=&T\sum_{n=-\infty}^\infty \Tr\left(\varepsilon+\frac12\zeta_n\frac{\partial
\varepsilon}{\partial\zeta_n}-\frac1{2\zeta_n}\bm{\nabla}\times\frac1{\mu^2}
\frac{\partial\mu}{\partial\zeta_n}\bm{\nabla}\times\right)\bm{\Gamma},
\end{eqnarray}
At zero temperature, this reduces to the expected general formula for a dispersive 
medium~\cite{Milton:2010yw,Candelas:1981qw}, where $\zeta=-i\omega$ is the imaginary 
frequency, 
\begin{equation}
E=U(T=0)=\int_{-\infty}^\infty \frac{d\zeta}{2\pi}\left[
\Tr\varepsilon\bm{\Gamma}+\frac12\zeta
\Tr \bm{\Gamma}\frac\partial{\partial\zeta}\varepsilon
-\frac12\frac1{\zeta}\bm{\nabla}\times\frac1{\mu^2}\frac{\partial\mu}
{\partial\zeta}\bm{\nabla}\times\bm{\Gamma}\right],
\end{equation}
which is what would be obtained by integrating the dispersive form of the energy density 
\cite{Schwinger:1998cla}, 
\begin{equation}
u(\mathbf{r})=
\frac12\left(\frac{d}{d\zeta}(\zeta\varepsilon)E^2+\frac{d}{d\zeta}
(\zeta\mu)H^2\right).
\end{equation}
For the case of anisotropic permittivity and permeability, provided the corresponding 
tensors are invertible, the same steps, starting from either the variational approach or 
from the electromagnetic energy density, lead to the following expression for the internal 
energy, 
\begin{equation}
U=T\sum_{n=-\infty}^\infty \Tr \left[\bm{\varepsilon}\cdot\bm{ \Gamma}+\frac12
\zeta_n \frac{d\bm{\varepsilon}}{d\zeta_n}\cdot\bm{\Gamma}
+\frac1{2\zeta_n}
\bm{\nabla}\times\frac{d\bm{\mu}^{-1}}{d\zeta_n}
\cdot\bm{\nabla}\times\bm{\Gamma}\right] \label{ie}
\end{equation}
that has a readily applicable form for calculating the Casimir self-energies of single 
objects, while the Casimir interaction energy between two objects is more conveniently 
evaluated using the multiple scattering expression~\cite{Kenneth:2006vr}. It is consistent 
with the variational statement, at zero temperature, 
\begin{equation}
\delta E=\int_{-\infty}^\infty \frac{d\zeta}{2\pi} \frac\zeta2
\frac{d}{d\zeta}\Tr
\left[\delta\bm{\varepsilon}\cdot\bm{\Gamma}-\delta\bm{\mu}^{-1}
\frac1{\zeta^2}\bm{\nabla\times\Gamma\times\overleftarrow{\nabla}'}\right].
\end{equation}
From this, it is quite direct to obtain the Lifshitz formula for parallel dielectric 
plates, as was done in Ref.~\cite{Schwinger:1977pa} for the pure permittivity case. The 
above discussion may not apply in the case of dissipation, see 
Refs.~\cite{Ginzburg:1989aetp,Brevik:2008ry}. 

Henceforth we specialize to the case of zero temperature. In this paper, we will be 
primarily considering self-energies in addition to the interaction energy, so regulation 
of integrals is necessary. Then, if we use point splitting in both time and space, the 
Casimir energy less the bulk (empty space) contribution, is 
\begin{eqnarray}
E-E_0&=&-\frac12\int_{-\infty}^\infty \frac{d\zeta}{2\pi}e^{i\zeta\tau}
\Tr\ln\bm{\Gamma}\,\bm{\Gamma}_0^{-1}\nonumber\\
&=&\frac12\int\frac{d\zeta}{2\pi}\frac{e^{i\zeta\tau}-1}{i\tau}\int
d^3x\Big\langle
\mathbf{x}\Big|\bm{\Gamma}^{-1}\frac{d\bm{\Gamma}}{d\zeta}-
\bm{\Gamma}_0^{-1}
\frac{d\bm{\Gamma}_0}{d\zeta}\Big|\mathbf{x}+\bm{\delta}\Big\rangle,
\end{eqnarray}
where $\tau$ and $\bm{\delta}$ are infinitesimal point-splitting parameters in time and 
space, to be taken to zero at the end of the calculation. In the second integral above, we 
have integrated by parts. Substituting, from Eq.~(\ref{degf}), 
\begin{equation}
\frac{d\bm{\Gamma}}{d\zeta}=-\bm{\Gamma}\frac{d\bm{\Gamma}^{-1}}{d\zeta}
\bm{\Gamma}=\bm{\Gamma}\frac{d\bm{\varepsilon}}{d\zeta}\bm{\Gamma}
+\bm{\Gamma}\frac1{\zeta^2}\bm{\nabla}\times\frac{d\bm{\mu}^{-1}}{d\zeta}
\bm{\nabla}\times\bm{\Gamma}+\frac2\zeta(\bm{\Gamma+\Gamma\varepsilon\Gamma}),
\end{equation}
Then, just as above, we obtain the zero-temperature, regulated form of the internal energy 
(\ref{ie}) 
\begin{equation}
E-E_0=\int_{-\infty}^\infty\frac{d\zeta}{2\pi}\frac{e^{i\zeta\tau}-1}{i\zeta
\tau}\Tr\left[\bm{\varepsilon\Gamma}+\frac\zeta2 \frac{d\bm{\varepsilon}}{d\zeta}
\bm{\Gamma}+\frac12\frac1{\zeta}\bm{\nabla}\times\frac{d\bm{\mu}^{-1}}
{d\zeta}\bm{\nabla}\times\bm{\Gamma}-\bm{\Gamma}_0\right],\label{genenform}
\end{equation}
where the trace includes a point-split integration over position.


\section{Electromagnetic $\delta$-function potential}
\label{d-potential}

The $\delta$-function potential model we use in this paper was introduced and 
extensively explored, for the planar geometry, in Ref.~\cite{Parashar:2012it}. 

An electromagnetic $\delta$-function potential describes an infinitesimally thin 
material with electric permittivity $\bm\varepsilon$ and magnetic permeability 
$\bm\mu$ defined in terms of a $\delta$-function, 
\begin{subequations}
\begin{eqnarray}
{\bm \varepsilon}({\bf x};\omega)&=&{\bf 1}+{\bm \lambda}_e(s;\omega)
\delta(s-s_0),
\\
{\bm \mu}({\bf x};\omega)&=&{\bf 1}+{\bm \lambda}_g(s;\omega)
\delta(s-s_0),
\end{eqnarray}
\end{subequations}
where $s$ represents the coordinate normal to the surface. We choose isotropic electric 
$\bm \lambda_e$ and magnetic $\bm \lambda_g$ susceptibilities of the material in the plane 
of the surface by requiring 
$\bm \lambda_e \equiv \text{diag}(\lambda_e^\perp,\lambda_e^\perp,\lambda_e^{||})$ and 
$\bm\lambda_g \equiv \text{diag}(\lambda_g^\perp,\lambda_g^\perp,\lambda_g^{||})$. 
The choice of isotropy in the plane of the surface ensures the separation of transverse 
electric (TE) and transverse magnetic (TM) modes.

In Ref.~\cite{Parashar:2012it}, we derived the conditions on the electric and magnetic 
fields at the boundary of such a material starting from the first order Maxwell's equations. 
We showed that a consistent set of boundary conditions on the fields only included the 
properties of the materials confined to the surface (shown below for a spherical 
$\delta$-function surface). Additional constraints on the components of the material 
properties transverse to the surface $\lambda^{||}$ were obtained from Maxwell's equations 
that lead to a necessarily anisotropic nature of the electromagnetic properties for 
materials described by a $\delta$-function potential. Specifically, we found $\lambda_e^{||}=0$ and 
$\lambda_g^{||}=0$. One must consider these discussions in light of 
Refs.~\cite{Barton:2013tp,Bordag:2014tp}. The $\lambda^{||}$ components do not appear in 
the boundary conditions. However, releasing the aforementioned conditions would require 
overconstraining the electric and magnetic fields according to Maxwell's equations. We shall 
extend this discussion further in Sec.~\ref{con-sph}.


\section{Electromagnetic $\delta$-function sphere}
\label{delta-sphere}
\begin{figure}
\includegraphics[width=5cm]{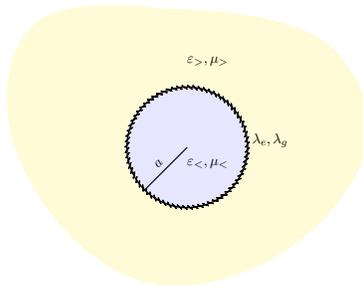}
\caption{A $\delta$-function sphere described by electric and magnetic couplings, 
$\lambda_e$ and $\lambda_g$, at the interface of two spherically symmetric media.}
\label{single-d123-sp}
\end{figure}

Consider an infinitesimally thin spherical shell at the interface of two spherically 
symmetric media, as shown in Fig.~\ref{single-d123-sp}. 
The electric permittivity and magnetic permeability for this is of the form
\begin{subequations}
\begin{eqnarray}
{\bm\varepsilon}(r) &=& \varepsilon^\perp(r)
\bm1_\perp + \varepsilon^{||}(r) \,\hat{\bm r} \hat{\bm r}, \\
{\bm\mu}(r) &=& \mu^\perp(r)
\bm1_\perp
+ \mu^{||}(r) \,\hat{\bm r} \hat{\bm r},
\end{eqnarray}%
\label{epmu-def}%
\end{subequations}
where
\begin{subequations}
\begin{eqnarray}
\varepsilon^{\perp,{||}}(r) &=& 1 + 
(\varepsilon^{\perp,{||}}_< -1) \,\theta(a-r)
+ (\varepsilon^{\perp,{||}}_> -1) \,\theta(r-a)
+ \lambda^{\perp,{||}}_e \,\delta (r-a), \\
\mu^{\perp,{||}}(r) &=& 1 + 
(\mu^{\perp,{||}}_< -1) \,\theta(a-r)
+ (\mu^{\perp,{||}}_> -1) \,\theta(r-a)
+ \lambda^{\perp,{||}}_g \,\delta (r-a).
\end{eqnarray}%
\label{eps-def}%
\end{subequations}
Here $\perp$ and $||$ refer to perpendicular and parallel to the radial direction 
$\hat{\bm  r}$ (which defines the direction of the surface vector at each point on 
the sphere).

In Heaviside-Lorentz units, the monochromatic components [proportional to 
$\exp(-i\omega t)$] of Maxwell's equations in the absence of charges and currents are 
\begin{subequations}
\begin{eqnarray}
{\bm \nabla} \times {\bf E} &=& i \omega {\bf B}, \label{MEcrossE} \\
-{\bm \nabla} \times {\bf H} &=& i \omega ({\bf D} + {\bf P}), \label{MEcrossB}
\end{eqnarray}%
\label{ME-cross}%
\end{subequations}
which imply ${\bm \nabla}\cdot{\bf B}=0$, and ${\bm \nabla}\cdot({\bf D}+ {\bf P})=0$, 
where ${\bf P}$ is an external source of polarization.

In the following we assume that the fields ${\bf D}$ and ${\bf B}$ are linearly dependent 
on the electric and magnetic fields ${\bf E}$ and ${\bf H}$ as
\begin{subequations}
\begin{eqnarray}
{\bf D}({\bf x},\omega) 
&=& {\bm \varepsilon}({\bf x};\omega) \cdot {\bf E}({\bf x},\omega), \\
{\bf B}({\bf x},\omega) 
&=& {\bm \mu}({\bf x};\omega) \cdot {\bf H}({\bf x},\omega).
\end{eqnarray}%
\label{DB=emuEB}%
\end{subequations} 
A vector field can be decomposed in the basis of the vector spherical harmonics as
\begin{equation}
{\bf V}({\bf r})=\sum_{lm}{\text V}^{(i)}_{lm}(r){\bf X}_{lm}^{(i)}(\theta,\phi), 
\label{VSH}
\end{equation}
where $i=1,2,r$ and 
${\bf X}_{lm}^{(i)}\equiv({\bf \Psi}_{lm}(\theta,\phi),{\bf \Phi}_{lm}(\theta,\phi),{\bf Y}_{lm}(\theta,\phi))$ are the basis vectors~\cite{Barrera:1985vsh}: 
\begin{eqnarray*}
{\bf \Psi}_{lm}(\theta,\phi) &=& \frac{r}{\sqrt{l(l+1)}}
                                 \bm\nabla\,\mathrm{Y}_{lm}(\theta,\phi),\\
{\bf \Phi}_{lm}(\theta,\phi) &=& \frac{r}{\sqrt{l(l+1)}}
                                 \hat{\bf r}\times\bm\nabla\,\text{Y}_{lm}(\theta,\phi),\\
{\bf Y}_{lm}(\theta,\phi)  &=& \hat{\bf r} \mathrm{Y}_{lm}(\theta,\phi) .
\end{eqnarray*}

Maxwell's equations in Eqs.~(\ref{ME-cross}) thus decouple into two modes: the transverse 
magnetic mode (TM) involves the field components $(E^{(1)}, H^{(2)}, E^{(r)})$,
\begin{subequations}
\begin{eqnarray}
\frac1r\frac{\partial}{\partial r}r E^{(1)}_{lm}(r)
&=& \frac{\sqrt{l(l+1)}}{r}E^{(r)}_{lm}(r)+ i\omega B^{(2)}_{lm}(r), 
\label{TM-first-E1} \\
\frac1r\frac{\partial}{\partial r}r H^{(2)}_{lm}(r)
&=& i\omega \left[D^{(1)}_{lm}(r) + P^{(1)}_{lm}(r)\right],
\label{TM-first-d-H2} \\
\frac{\sqrt{l(l+1)}}{r}H^{(2)}_{lm}(r) &=& i\omega \left[D^{(r)}_{lm}(r) + P^{(r)}_{lm}(r)\right],
\label{TM-first-H2}%
\end{eqnarray}%
\label{TM-first-order}%
\end{subequations}
and the transverse electric mode (TE) involves the field components 
$(H^{(1)}, E^{(2)}, H^{(r)})$, 
\begin{subequations}
\begin{eqnarray}
\frac1r\frac{\partial}{\partial r}r H^{(1)}_{lm}(r)
&=& \frac{\sqrt{l(l+1)}}{r}H^{(r)}_{lm}(r)-i\omega \left[D^{(2)}_{lm}(r)+P^{(2)}_{lm}(r)\right] , 
\label{TE-first-H1} \\
\frac1r\frac{\partial}{\partial r}r E^{(2)}_{lm}(r)
&=& -i\omega B^{(1)}_{lm}(r)\label{TE-first-d-E2},\\
\frac{\sqrt{l(l+1)}}{r}E^{(2)}_{lm}(z) &=& -i\omega B^{(r)}(r).
\label{TE-first-E2} %
\end{eqnarray}%
\label{TE-first-order}%
\end{subequations}

\subsection{Boundary conditions}
\label{boundary-c}

The boundary conditions on the electric and magnetic fields ${\bf E}$ and ${\bf H}$ are 
obtained by integrating across the $\delta$-function boundary. We get additional 
contributions to the standard boundary conditions at the interface of two media due to the 
presence of the $\delta$-function sphere. The only requirement on the the electric field 
${\bf E}$ and magnetic field ${\bf H}$ is that they are free from any $\delta$-function 
type singularities, which is evident from the second-order differential equation of the 
fields. The boundary conditions on the fields are
%
\begin{subequations}
\begin{align}
&\text{\underline{TM}}&\text{\underline{TE}}&\nonumber\\
E^{(1)}_{lm}(r)\Big|^{r=a+}_{r=a-} &= i\omega \lambda_g^\perp H^{(2)}_{lm}(a),
& H^{(1)}_{lm}(r)\Big|^{r=a+}_{r=a-} &= -i\omega \lambda_e^\perp E^{(2)}_{lm}(a),\\
H^{(2)}_{lm}(r)\Big|^{r=a+}_{r=a-} &= i\omega \lambda_e^\perp E^{(1)}_{lm}(a),
& E^{(2)}_{lm}(r)\Big|^{r=a+}_{r=a-} &= -i\omega \lambda_g^\perp H^{(1)}_{lm}(a)\\
D^{(r)}_{lm}(r)\Big|^{r=a+}_{r=a-} &= \frac{\sqrt{l(l+1)}}{a}  \lambda_e^\perp E^{(1)}_{lm}(a),
& B^{(r)}_{lm}(r)\Big|^{z=a+}_{z=a-} &= \frac{\sqrt{l(l+1)}}{a} \lambda_g^\perp H^{(1)}_{lm}(a) \label{TM-bc-D3}.%
\end{align}%
\label{field-bc}%
\end{subequations}
We evaluate quantities that are discontinuous on the $\delta$-function sphere 
using the averaging prescription, introduced earlier in 
Refs.~\cite{Cavero-Pelaez:2008ncg1,Cavero-Pelaez:2008ncg2}. In addition we get the constraints 
\begin{equation}
\lambda_e^{||} E^{(r)}(a) = 0 
\qquad \text{and} \qquad
\lambda_g^{||} H^{(r)}(a) = 0,
\label{constraint}%
\end{equation}
which implies that optical properties of the magneto-electric $\delta$-function sphere are 
necessarily anisotropic unless $E^{(r)}_{lm}(a)=0$ and $H^{(r)}_{lm}=0$. The constraints in 
(\ref{constraint}) 
are not obvious from the second-order equations for the fields. However, they will appear 
in the same form if we try to obtain boundary conditions on $E^{(r)}_{lm},H^{(r)}_{lm}$ 
from their respective second order differential equations upon integration. If we do not 
take into account of these constraints, it may appear that $\lambda^{||}_{e,g}$ have 
consequences on the optical properties of the $\delta$-sphere. (See discussion in 
Sec.~\ref{con-sph}.)

The Maxwell equations in Eqs.~(\ref{TM-first-order}) and (\ref{TE-first-order}), which are 
in first order form, can be combined to yield the second order differential equations 
[with ${\bm\varepsilon}=\text{diag}(\varepsilon^\perp,\varepsilon^\perp,\varepsilon^{||})$ 
and ${\bm\mu}= \text{diag}(\mu^\perp,\mu^\perp,\mu^{||})$]
\begin{subequations}
\begin{eqnarray}
\left[ - \frac{\partial}{\partial r} \frac{1}{\varepsilon^\perp(r)}
\frac{\partial}{\partial r} + \frac{1}{\varepsilon^{||}(z)} \frac{l(l+1)}{r^2}
-\omega^2 \mu^\perp(r) \right] r H^{(2)}_{lm}(r) 
&=&- i\omega \frac{\partial}{\partial r} r\frac{P^{1)}_{lm}(r)}{\varepsilon^\perp(r)}
+i \omega \sqrt{l(l+1)} \frac{P^{(r)}_{lm}(r)}{\varepsilon^{||}(r)},
\label{2order-D3} \\
\left[ - \frac{\partial}{\partial r} \frac{1}{\mu^\perp(r)}
\frac{\partial}{\partial r} + \frac{1}{\mu^{||}(z)} \frac{l(l+1)}{r^2}
-\omega^2 \varepsilon^\perp(r) \right] rE^{(2)}_{lm}(r)  &=& \omega^2 P^{(2)}_{lm}(r) .
\label{2order-B3}%
\end{eqnarray}%
\label{2order-D3B3}%
\end{subequations}
The remaining field components can be expressed in terms of 
$H^{(2)}_{lm}(r)$ and $E^{(2)}_{lm}(r)$.

\subsection{Green's dyadics}
\label{green-dyad}

We use the Green's function technique to obtain the electric and magnetic fields 
${\bf E}({\bf x};i\omega)$ and ${\bf H}({\bf x};i\omega)$: 
\begin{equation}
{\bf E}({\bf x};i\omega) 
= \int d^3x^\prime \,{\bm \Gamma} ({\bf x},{\bf x}^\prime;i\omega) \cdot 
  {\bf P}({\bf x}^\prime;i\omega)
\qquad \text{and} \qquad
{\bf H}({\bf x};i\omega) 
= \int d^3x^\prime \,{\bm \Phi} ({\bf x},{\bf x}^\prime;i\omega) \cdot 
  {\bf P}({\bf x}^\prime;i\omega),
\label{E=Gp,H=PP}%
\end{equation}
in terms of the electric Green's dyadic ${\bm \Gamma} ({\bf x},{\bf x}^\prime)$ and 
magnetic Green's dyadic ${\bm \Phi} ({\bf x},{\bf x}^\prime)$ respectively. (We have used 
the same notation ${\bm \Phi} ({\bf x},{\bf x}^\prime)$ as one of the basis vectors of 
the vector spherical harmonics, but the two have different arguments.) Green's 
dyadics can be expanded in terms of the vector spherical harmonics as
\begin{subequations}
\begin{eqnarray}
{\bm \Gamma} ({\bf x},{\bf x}^\prime)
&=&\sum_{lm}{\bf X}^{\text{T}}_{lm}(\theta,\phi)
{\bm \gamma}_l (r,r^\prime){\bf X}^{*}_{lm}(\theta,\phi),\\
{\bm \Phi} ({\bf x},{\bf x}^\prime)
&=&\sum_{lm}{\bf X}^{\text{T}}_{lm}(\theta,\phi)
{\bm \phi}_l (r,r^\prime){\bf X}^{*}_{lm}(\theta,\phi),
\label{dyadic-VSH}
\end{eqnarray}
\end{subequations}
where ${\bf X}^{\text{T}}_{lm}(\theta,\phi)$ is the transpose and 
${\bf X}^{*}_{lm}(\theta,\phi)$ is the complex conjugate of the basis vector. The reduced 
Green's matrices ${\bm \gamma}_l (r,r^\prime)$ and ${\bm \phi}_l (r,r^\prime)$ can be 
solved in terms of scalar Green's functions $g^H_l(r,r^\prime)$ and $g^E_l(r,r^\prime)$
\begin{equation}
{\bm\gamma}_l
= \begin{array}{c|c}
(1) \hspace{20mm} (2) \hspace{20mm} (r) & \\[0mm] \hline \\[-3mm]
\left[ \begin{array}{ccc}
\frac{1}{\varepsilon^\perp}\frac1r \frac{\partial}{\partial r}
\frac{1}{\varepsilon^{\prime\perp}}\frac{1}{r^\prime} \frac{\partial}{\partial r^\prime}
 r^\prime g^H_l & 0 &
\frac{\sqrt{l(l+1)}}{r^\prime}\frac{1}{\varepsilon^{\prime||}}
\frac{1}{\varepsilon^\perp} \frac1r\frac{\partial}{\partial r} r g^H_l \\[2mm]
0 & \omega^2 g^E_l & 0 \\[2mm]
\frac{\sqrt{l(l+1)}}{r}\frac{1}{\varepsilon^{||}}
\frac{1}{\varepsilon^{\prime\perp}} \frac{1}{r^\prime}\frac{\partial}{\partial r^\prime} 
r^\prime g^H_l & 
0 &\frac{l(l+1)}{rr^\prime}\frac{1}{\varepsilon^{||}}
\frac{1}{\varepsilon^{\prime||}} g^H_l
\end{array} \right]
& \begin{array}{c} (1) \\[3mm] (2) \\[3mm] (r) \end{array}
\end{array}
\label{Gamma=gE}
\end{equation}
and
\begin{equation}
{\bm\phi_l}
= i\omega \left[ \begin{array}{ccc}
0 & \frac{1}{\mu^\perp} \frac1r\frac{\partial}{\partial r}r g^E_l
& 0 \\[2mm]
\frac{1}{\varepsilon^{\prime\perp}}
\frac1r\frac{\partial}{\partial r^\prime}r^\prime g^H_l & 0 &
\frac{\sqrt{l(l+1)}}{r^\prime}\frac{1}{\varepsilon^{\prime||}} g^H_l \\[2mm]
0 & \frac{\sqrt{l(l+1)}}{r}\frac{1}{\mu^{||}} g^E & 0
\end{array} \right],
\label{Phi=gH}
\end{equation}
where we have suppressed the $r$ and $r^\prime$ dependence and, $\varepsilon^\prime$ is 
$\varepsilon(r^\prime)$. In Eq.~(\ref{Gamma=gE}) we have omitted a contact term 
involving $\delta(r-r^\prime)$, 
\begin{equation*}
-\frac{1}{r r^\prime}\left[\begin{array}{ccc}
\frac{\delta(r-r^\prime)}{\varepsilon^\perp(r^\prime)}& 0 & 0)\\[2mm]
0 & 0 & \\[2mm]
0 & 0 & \frac{\delta(r-r^\prime)}{\varepsilon^{||}(r^\prime)}
\end{array}\right].
\end{equation*}
The magnetic Green's function $g^H_l(r,r^\prime)$ and the electric 
Green's function $g^E_l(r,r^\prime)$ satisfy 
\begin{subequations}
\begin{eqnarray}
\left[ - \frac1r\frac{\partial}{\partial r}r \frac{1}{\varepsilon^\perp(r)}
\frac1r\frac{\partial}{\partial r} r + \frac{\l(l+1)}{r^2}\frac{1}
{\varepsilon^{||}(z)} 
-\omega^2 \mu^\perp(z) \right] g^H_l(r,r^\prime) &=& \frac{\delta(r-r^\prime)}{r^2}, 
\label{greenH} \\
\left[ - \frac1r\frac{\partial}{\partial r}r \frac{1}{\mu^\perp(r)}
\frac1r\frac{\partial}{\partial r} r + \frac{l(l+1)}{r^2}\frac{1}
{\mu^{||}(z)} 
-\omega^2 \mu^\perp(z) \right] g^E_l(r,r^\prime)  &=& \frac{\delta(r-r^\prime)}{r^2}, 
\label{greenE}
\end{eqnarray}%
\label{green-funs}%
\end{subequations}
where the material properties $\varepsilon^\perp(r)$ and $\mu^\perp(r)$ are defined in 
Eqs.~(\ref{eps-def}). 

\subsection{Magnetic and electric Green's functions}
\label{green-fn}

We obtain the boundary conditions on the magnetic Green's functions using 
Eqs.~(\ref{field-bc}) for the TM mode,
\begin{subequations}
\begin{eqnarray}
g^H_l \Big|^{r=a+}_{r=a-}
&=& \lambda^\perp_e \frac{1}{\varepsilon^\perp} \frac1r
\frac{\partial}{\partial r}
r g^H_l \bigg|_{r=a} , \\
\frac{1}{\varepsilon^\perp} \frac1r
\frac{\partial}{\partial r}r
g^H_l \bigg|^{r=a+}_{r=a-}
&=& \zeta^2 \lambda^\perp_g g^H_l\Big|_{r=a}.
\end{eqnarray}%
\label{gH-bc}%
\end{subequations} 
Similarly, using Eqs.~(\ref{field-bc}) for TE mode, the boundary conditions on the 
electric Green's function are 
\begin{subequations}
\begin{eqnarray}
g^E_l \Big|^{r=a+}_{r=a-}
&=& \lambda^\perp_g \frac{1}{\mu^\perp} \frac1r
\frac{\partial}{\partial r}r
g^E_l \bigg|_{r=a}, \\
\frac{1}{\mu^\perp} \frac1r
\frac{\partial}{\partial r}r
g^E_l \bigg|^{r=a+}_{r=a-}
&=& \zeta^2 \lambda^\perp_e g^E_l\Big|_{r=a}.
\end{eqnarray}%
\label{gE-bc}%
\end{subequations} 
Here $\zeta=-i\omega$ is the imaginary frequency obtained after a Euclidean rotation. 

\begin{figure}
\includegraphics[width=50mm]{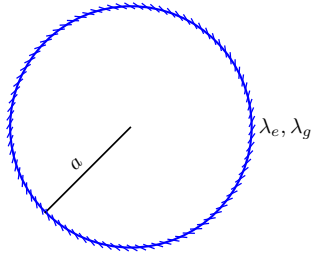}
\caption{A $\delta$-function sphere in vacuum.}
\label{single-d-sp-vac}
\end{figure}

The general solution for the spherical magnetic scalar Green's function for a system 
shown in Fig.~\ref{single-d123-sp} is given in Ref.~\cite{Shajesh:2017ax}. In this paper, 
we are particularly interested in the case when the media surrounding the 
$\delta$-function sphere is vacuum, as shown in Fig.~\ref{single-d-sp-vac}. 
In this case, the magnetic scalar Green's function is 
\begin{equation}
g^H_l(r,r^\prime;i\zeta) = \begin{cases}
\frac{2}{\pi} \zeta \left[\text{i}_l(\zeta r_<) \text{k}_l(\zeta r_>) 
+ \sigma^\text{scatt}_{l,<>} \, \text{i}_l(\zeta r) \text{i}_l(\zeta r^\prime) 
\right], \quad & r,r^\prime < a, \\[3mm]
\frac{2}{\pi} \zeta \left[\text{i}_l(\zeta r_<) \text{k}_l(\zeta r_>) 
+ \sigma^\text{scatt}_{l,><} \,
\text{k}_l(\zeta r) \text{k}_l(\zeta r^\prime) 
\right], & a<r,r^\prime, \\[3mm]
\frac{2}{\pi} \zeta \,\sigma^\text{abs}_{l,><} \,
\text{i}_l(\zeta r) \text{k}_l(\zeta r^\prime),
& r<a<r^\prime, \\[3mm]
\frac{2}{\pi} \zeta \,\sigma^\text{abs}_{l,<>} \,
\text{k}_l(\zeta r) \text{i}_l(\zeta r^\prime),
& r^\prime<a<r,
\end{cases}
\label{gH-sol}
\end{equation}
where the scattering coefficient $\sigma^\text{scatt}$ and absorption coefficients 
$\sigma^\text{abs}$ are
\begin{subequations}
\label{scatt-coeff}
\begin{eqnarray}
\sigma^\text{scatt}_{l,<>} &=& 
- \frac{ \left[ \zeta \lambda^\perp_e \, \bar{\text{k}} \bar{\text{k}}
- \zeta \lambda^\perp_g \, \text{k} \text{k} \right] }
{ \left[ (1+\zeta^2 \lambda^\perp_e \lambda^\perp_g/4)
(\text{i} \bar{\text{k}} - \bar{\text{i}} \text{k})
+ \zeta\lambda^\perp_e \, \bar{\text{i}} \bar{\text{k}} 
- \zeta\lambda^\perp_g \, \text{i} \text{k} \right] }
\xrightarrow[\lambda^\perp_g\to\infty]{\lambda^\perp_e\to\infty} 0, \\
\sigma^\text{scatt}_{l,><} &=& 
- \frac{ \left[ \zeta \lambda^\perp_e \, \bar{\text{i}} \bar{\text{i}}
- \zeta \lambda^\perp_g \, \text{i} \text{i} \right] }
{ \left[ (1+\zeta^2 \lambda^\perp_e \lambda^\perp_g/4)
(\text{i} \bar{\text{k}} - \bar{\text{i}} \text{k})
+ \zeta\lambda^\perp_e \, \bar{\text{i}} \bar{\text{k}} 
- \zeta\lambda^\perp_g \, \text{i} \text{k} \right] }
\xrightarrow[\lambda^\perp_g\to\infty]{\lambda^\perp_e\to\infty} 0, \\
\sigma^\text{abs}_{l,><} =
\sigma^\text{abs}_{l,<>} &=&
\frac{ (1+\zeta^2 \lambda^\perp_e \lambda^\perp_g/4)
(\text{i} \bar{\text{k}} - \bar{\text{i}} \text{k}) }
{ \left[ (1+\zeta^2 \lambda^\perp_e \lambda^\perp_g/4)
(\text{i} \bar{\text{k}} - \bar{\text{i}} \text{k})
+ \zeta\lambda^\perp_e \, \bar{\text{i}} \bar{\text{k}} 
- \zeta\lambda^\perp_g \, \text{i} \text{k} \right] }
\xrightarrow[\lambda^\perp_g\to\infty]{\lambda^\perp_e\to\infty} -1,
\end{eqnarray}
\end{subequations}
where we have suppressed the argument and subscript to save typographical space. We use 
the modified spherical Bessel functions 
$\text{i}_l(t)$ and $\text{k}_l(t)$~\cite{NIST:2010fm} that are related to the modified 
Bessel functions as 
\begin{subequations}
\begin{eqnarray}
\text{i}_l(t) &=& \sqrt{\frac{\pi}{2t}} I_{l+\frac{1}{2}}(t), \\
\text{k}_l(t) &=& \sqrt{\frac{\pi}{2t}} K_{l+\frac{1}{2}}(t).
\end{eqnarray}%
\end{subequations}
In particular $\text{i}_l(t) = \text{i}^{(1)}_l(t)$, which is the modified spherical 
Bessel function of the first kind, and together with $\text{k}_l(t)$ are a satisfactory 
pair of solutions in the right half of the complex plane. We have also used bars to define 
the following operations on the modified spherical Bessel functions
\begin{eqnarray}
\bar{\text{i}}_l(t) 
&=& \bigg( \frac{1}{t} + \frac{\partial}{\partial t} \bigg) \text{i}_l(t), \\
\bar{\text{k}}_l(t) 
&=& \bigg( \frac{1}{t} + \frac{\partial}{\partial t} \bigg) \text{k}_l(t).
\end{eqnarray}
Solutions for the electric Green's function can be obtained from the magnetic 
Green's function by replacing ${\bm \varepsilon} \leftrightarrow {\bm \mu}$ and $H\to E$.

We show the values of coefficients corresponding to a perfectly conducting electric and 
magnetic spherical shell in the rightmost listing in Eq. (\ref{scatt-coeff}). Notice that 
the spherical shell becomes completely transparent in this extreme limit, and the total 
transmission is accompanied by a phase change of $\pi$.

It is crucial to emphasize the fact that even though we explicitly considered materials 
with $\lambda^{||}_e$ and  $\lambda^{||}_g$  in Eqs.~(\ref{eps-def}), the solutions to the 
Green's functions of Eq. (\ref{gH-sol}) are independent of $\lambda^{||}_e$ and 
$\lambda^{||}_g$ because the boundary conditions in (\ref{gE-bc}) do not depend on the 
parallel components of the coupling. The Green's functions of Eqs. (\ref{gH-sol}) 
determine the fields unambiguously everywhere except on the $\delta$-function plate, where 
we use an averaging prescription. The implication is that there are no observable 
consequences of $\lambda^{||}_e$ and $\lambda^{||}_g$. 


\section{Casimir interaction energy between two concentric electric $\delta$-function 
spheres}
\label{con-sph}

\begin{figure}
\includegraphics[width=45mm]{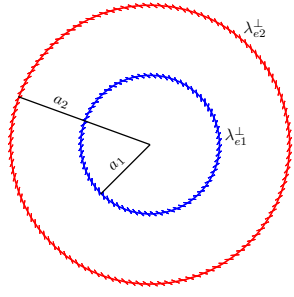}
\caption{Concentric $\delta$-function spheres with purely electric material properties.}
\label{con-d-sp}
\end{figure}

As a check for the formalism developed for a $\delta$-function sphere, we first calculate 
the Casimir interaction energy between two concentric electric $\delta$-function spheres 
as shown in Fig.~\ref{con-d-sp}, where $\lambda_g=0$. We set $\lambda^{||}=0$ to satisfy 
the constraint given in Eq.~(\ref{constraint}). In the asymptotic flat plate limit, i.e. 
small angle and large radius (See Fig.~\ref{Flat-pLimit}), the interaction energy between 
the  concentric $\delta$-function spheres should reproduce the interaction energy between 
two $\delta$-function plates. This limit coincides with the PFA for the spherical surfaces.

The Casimir interaction energy between two concentric $\delta$-function spheres is
\begin{equation}
E_{12}=\frac12\int_{-\infty}^{\infty}\frac{d\zeta}{2\pi}\sum_{l=1}^\infty
\sum_{m=-l}^l{\Tr}\,\ln\left(\bf 1-{\bf K }\right),
\label{E-int}
\end{equation} 
where the notation $\Tr$ implies trace on both space coordinates ($\Tr_s$) and matrix 
coordinates ($\tr$). The kernel ${\bf K}$ is
\begin{equation}
{\bf K}(r,r^\prime)={\bm \gamma}_{1}(r,r^\prime)\cdot{\bm \lambda}_{e1}^\perp(r^\prime)
\delta(r^\prime-a_1)
             \cdot{\bm \gamma}_{2}(r^\prime,r)\cdot{\bm \lambda}_{e2}^\perp(r)
\delta(r-a_2).
\end{equation}
The interaction energy between two non-overlapping objects is always finite, hence we have 
dropped the cut-off parameters. Further, we can decompose Eq.~(\ref{E-int}) into the TE 
and TM parts using the identity $\tr\,\ln {\bf K}=\ln\, \text{det}\,{\bf K}$
\begin{equation}
{E}_{12}=
 \frac12\int_{-\infty}^{\infty}\frac{d\zeta}{2\pi}\sum_{l}(2l+1)\int d^3 x 
\left[\ln\left(1-K^E\right)+ \tr \ln \left(\bf 1 - {\bf K}^H\right)\right],
\end{equation}
where $\text{K}^E$ corresponds to the 22 component of the kernel, which depends on the 
22 component of Green's dyadic given in (\ref{Gamma=gE}), and ${\bf K}^H$ is rest of 
the matrix. The sum on $m$ is trivial as the magnetic Green's function in Eq. 
(\ref{gH-sol}) and the corresponding electric Green's function are independent of $m$. 
One can verify that $\text{det}\,{\bf K}^H=0$, which implies that
\begin{equation}
\Tr \,\ln \left(1 - {\bf K}^H\right) =\ln \left(1 - \Tr\,{\bf K}^H\right).
\end{equation}
The coefficients in Eq.~(\ref{scatt-coeff}) for this case take the form
\begin{subequations}
\begin{eqnarray}
\sigma^\text{scatt}_{l,<>} &=& 
- \frac{\zeta \lambda^\perp_e \, \bar{\text{k}} \bar{\text{k}}}
{ \left[ (\text{i} \bar{\text{k}} - \bar{\text{i}} \text{k})
+ \zeta\lambda^\perp_e \, \bar{\text{i}} \bar{\text{k}} \right] }
\xrightarrow{\lambda^\perp_e\to\infty}
- \frac{\bar{\text{k}}}{\bar{\text{i}}}, \\
\sigma^\text{scatt}_{l,><} &=& 
- \frac{\zeta \lambda^\perp_e \, \bar{\text{i}} \bar{\text{i}}}
{ \left[ (\text{i} \bar{\text{k}} - \bar{\text{i}} \text{k})
+ \zeta\lambda^\perp_e \, \bar{\text{i}} \bar{\text{k}} \right] }
\xrightarrow{\lambda^\perp_e\to\infty}
- \frac{\bar{\text{i}}}{\bar{\text{k}}}, \\
\sigma^\text{abs}_{l,><} =
\sigma^\text{abs}_{l,<>} &=&
 \frac{(\text{i} \bar{\text{k}} - \bar{\text{i}} \text{k})}
{ \left[ (\text{i} \bar{\text{k}} - \bar{\text{i}} \text{k})
+ \zeta\lambda^\perp_e \, \bar{\text{i}} \bar{\text{k}} \right] }
\xrightarrow{\lambda^\perp_e\to\infty} 0,
\end{eqnarray}
\end{subequations}
where the rightmost values are given for a perfectly conducting spherical shell. The TE 
part of the interaction energy between concentric $\delta$-function spheres is 
\begin{equation}
E^\mathrm{TE}_{12}=\frac12\int_{-\infty}^{\infty}\frac{d\zeta}{2\pi}\sum_{l}(2l+1)
\ln\left[1-\frac{\zeta \lambda_{e1}^\perp}{\frac{\pi}{2(\zeta a_1)^2}
                  +\zeta \lambda_{e1}^\perp{\text i}_l(\zeta a_1){\text k}_l(\zeta a_1)}
                 \frac{\zeta \lambda_{e2}^\perp}{\frac{\pi}{2(\zeta a_2)^2}
                  +\zeta \lambda_{e2}^\perp{\text i}_l(\zeta a_2){\text k}_l(\zeta a_2)}
                  {\text i}_l^2(\zeta a_1) {\text k}_l^2(\zeta a_2)\right]
\end{equation}
and the TM part of the interaction energy is 
\begin{equation}
E^\mathrm{TM}_{12}=\frac12\int_{-\infty}^{\infty}\frac{d\zeta}{2\pi}\sum_{l}(2l+1)
\ln\left[1-\frac{\zeta \lambda_{e1}^\perp}{\frac{\pi}{2(\zeta a_1)^2}
    -\zeta \lambda_{e1}^\perp {\bar {\text i}}_l(\zeta a_1){\bar{\text k}}_l(\zeta a_1)}
\frac{\zeta \lambda_{e2}^\perp}{\frac{\pi}{2(\zeta a_2)^2}
    -\zeta \lambda_{e2}^\perp {\bar {\text i}}_l(\zeta a_2){\bar{\text k}}_l(\zeta a_2)}
                  {\bar {\text i}}_l^2(\zeta a_1) {\bar {\text k}}_l^2(\zeta a_2)\right].
\end{equation}
In the asymptotic flat plate limit, which is equivalent to taking the uniform 
asymptotic expansion of the Bessel functions for $l \to \infty$, and  
keeping the distance between two spheres $a$ constant, we obtain the TE and TM mode 
interaction energies per unit area  $\cal{E}$ for two parallel $\delta$-function plates:
\begin{figure}
\includegraphics[width=40mm]{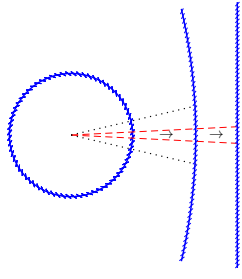}
\caption{Flat plate limit: For a very large radius and small angle approximation a 
spherical surface is locally flat.}
\label{Flat-pLimit}
\end{figure}%
\begin{subequations}
\begin{eqnarray}
{\cal E}^\mathrm{TE}_{12}&=&\frac12\int_{-\infty}^{\infty}\frac{d\zeta}{2\pi}\int_{-\infty}^{\infty}\frac{d^2k}{(2\pi)^2}
\ln\left(1-\frac{\lambda_{e1}^\perp}{\lambda_{e1}^\perp+\frac{2\kappa}{\zeta^2}} 
\frac{\lambda_{e2}^\perp}{\lambda_{e2}^\perp+\frac{2\kappa}{\zeta^2}}
e^{-2\kappa a}\right),
\label{TE-E-flat}
\\
{\cal E}^\mathrm{TM}_{12}&=&\frac12\int_{-\infty}^{\infty}\frac{d\zeta}{2\pi}\int_{-\infty}^{\infty}\frac{d^2k}{(2\pi)^2}
\ln\left(1-\frac{\lambda_{e1}^\perp}{\lambda_{e1}^\perp+\frac{2}{\kappa}} 
\frac{\lambda_{e2}^\perp}{\lambda_{e2}^\perp+\frac{2}{\kappa}}e^{-2\kappa a}\right),
\label{TM-E-flat}
\end{eqnarray}
\label{E-flat}
\end{subequations}
which gives the correct perfect conductor limit. 

It is worth discussing the implication of the choice $\lambda^{||}=0$ here. We had 
pointed out in Eq.~(\ref{constraint}) that a $\delta$-function boundary imposes constraints, 
$\lambda_e^{||} E^{(r)}(a) = 0$ and $\lambda_g^{||} H^{(r)}(a) = 0$. Additionally, the 
boundary conditions on the fields given by Eq.~(\ref{field-bc}) are independent of 
$\lambda^{||}$, thus the reflection coefficients appearing in Green's function are 
independent of $\lambda^{||}$. These observations suggested a necessarily anisotropic 
nature of the $\delta$-function material with $\lambda$. Based on the above observations, 
we calculated the Casimir interaction energies using the multiple scattering method in 
Eq.~(\ref{E-int}) for the TE and 
TM mode for $\bm \lambda=\left(\lambda^\perp,\lambda^\perp,0\right)$ requiring 
$\lambda^{||}=0$, which for the parallel plate case are given in Eq. (\ref{E-flat}). Let 
us explore the case when we ignore this constraint and keep $\lambda^{||}\neq 0$ 
in $\bm \lambda=\left(\lambda^\perp,\lambda^\perp,\lambda^{||}\right)$. In this case, the 
interaction energy of the TM mode would become
\begin{equation*}
\frac12\int_{-\infty}^{\infty}\frac{d\zeta}{2\pi}\int_{-\infty}^{\infty}\frac{d^2k}{(2\pi)^2}
\ln\left[1-\left(\frac{\lambda_{e1}^\perp}{\lambda_{e1}^\perp+\frac{2}{\kappa}}
                     +k^2_\perp\frac{\lambda_{e1}^{||}}{2\kappa} \right)
                     \left(\frac{\lambda_{e2}^\perp}{\lambda_{e2}^\perp+\frac{2}{\kappa}}
                     +k^2_\perp\frac{\lambda_{e2}^{||}}{2\kappa} \right)e^{-2\kappa a}
\right].
\end{equation*}
This would suggest the identification of a TM ``reflection coefficient'' of a single 
$\delta$-function plate of the form
\begin{equation}
\frac{\lambda_{ei}^\perp}{\lambda_{ei}^\perp+\frac{2}{\kappa}}
                     +k^2_\perp\frac{\lambda_{ei}^{||}}{2\kappa} ,
\label{rTM-wrong}
\end{equation}
which is inconsistent with the reflection coefficients found in the solutions of Green's 
functions using the boundary conditions as explained below.  
First we note that the second term is the added contribution due to the inclusion of the 
nonzero 
$\lambda^{||}$ in the Casimir interaction energy calculation. But this is not satisfactory 
because the reflection coefficient in Eq.~(\ref{rTM-wrong}) doesn't have a finite limit in 
$\lambda^{||}\to \infty$. At best, it suggests a weak behavior of $\lambda^{||}$. 
In other words, a $\delta$-function material can only have high conductivity in the 
surface of the material, which seems to be a physically viable option for an 
infinitesimally thin material. Secondly and more importantly, in Ref.~\cite{Parashar:2012it,*Shajesh:2017axs} we 
show by a direct calculation of Green's function for $\delta$-function plates that the 
reflection coefficients do not depend on the $\lambda^{||}$, which is a consequence of the 
fact that the boundary conditions are not contingent on the $\lambda^{||}$. 
Thus, the appearance of $\lambda^{||}$ in (\ref{rTM-wrong}) belies the adage that one can 
determine the Casimir (Lifshitz) interaction energy once the reflection coefficients are 
known. These observations strongly advocate for $\lambda^{||}=0$ as the consistent choice.


\section{Self energy of a $\delta$-function sphere}
\label{self-E-el}

We are particularly interested in analyzing the self-energy of a $\delta$-function sphere, 
which in general has divergent parts. We will use the point-splitting regulator for 
evaluating the self-energy. For a nonzero value of the point-splitting regulator 
$\bm\delta$ (where $\bm\delta$ has both temporal and spatial point-splitting components) 
the energy remains finite but it diverges in the limit $\bm\delta\to 0$. 
Using Maxwell's equations and the definition of Green's dyadic in Eq.~ (\ref{E=Gp,H=PP}), 
we can rewrite the bulk subtracted energy given in Eq.~(\ref{genenform}) for a dispersive 
magneto-electric material as 
\begin{equation}
E-E_0 =\frac12\int_{-\infty}^{\infty}\frac{d\zeta}{2\pi}\int d^3x\,
\frac{e^{i\zeta\tau}-1}{i\zeta\tau}\tr
\Big[2{\bm \varepsilon}\cdot{\bm \Gamma} 
+ \zeta\frac{d\bm \varepsilon}{d\zeta}
\cdot{\bm \Gamma} -{\bm \mu}^{-1}\cdot\frac{d\bm \mu}{d\zeta} 
\cdot({\bm \Phi}\times\overleftarrow{\bm \nabla^\prime})
-2{\bm \Gamma}_0 \Big]\bigg|_{{\bf x}^\prime={\bf x}+{\bm \delta}}.
\label{E-disp}
\end{equation}

Using the expansion of Green's dyadics in vector spherical harmonics and choosing the 
point-splitting in both 
temporal and spatial directions we can express the self-energy in terms of reduced 
Green's dyadic
\begin{subequations}
\begin{equation}
E-E_0=\frac12\int_{-\infty}^{\infty}\frac{d\zeta}{2\pi}
       \frac{e^{i\zeta\tau}-1}{i\zeta\tau}\sum_l(2l+1)P_l(\cos\delta)\,I ,
\end{equation}
where
\begin{equation}
I=\int_0^\infty r^2dr\,
	\tr\Big[2{\bm \varepsilon}\cdot{\bm \gamma}  + \zeta\frac{d\bm \varepsilon}{d\zeta}
	\cdot{\bm \gamma} -{\bm \mu}^{-1}\cdot\frac{d\bm \mu}{d\zeta}
	\cdot({\bm \phi}\times\overleftarrow{\bm \nabla^\prime})
-2{\bm \gamma}_0
\Big]\Big|_{r^\prime=r} . \label{I}     
\end{equation}%
\label{E-disp-sp}%
\end{subequations}%

In the following, we shall continue with the choice of purely electric $\delta$-function 
materials, i.e. $\lambda_g=0$. We also need to choose a particular model to define the 
frequency-dependent 
coupling constant in order to account for the dispersion, which is relevant for the energy 
calculated from Eq.~(\ref{E-disp}). A sort of plasma model, 
$\lambda^\perp_e=\zeta_p/\zeta^2$, is a straightforward choice, where $\zeta_p$ is an 
effective plasma frequency. (This is identical to Barton's hydrodynamical model, where the 
parameter $\zeta_p$ corresponds to the characteristic wavenumber \cite{Barton:2005eps}.)

\subsection{Self energy of an electric $\delta$-function plate}
\label{d-sph-el-self-E}

We first apply the energy expression in Eq.~(\ref{E-disp}) to calculate the self-energy of 
a purely electric $\delta$-function plate ($\lambda^\perp_g=0$ and $\lambda^\perp_e=\zeta_p/\zeta^2$). Keeping only the 
spatial cut-off $\bm\delta_\perp$ we obtain the energy per unit area,
\begin{equation}
{\cal E}-{\cal E}_0=\frac12\int_{-\infty}^{\infty}\frac{d\zeta}{2\pi}
                   \int\frac{d^2k}{(2\pi)^2}
          e^{i{\bf k}\cdot{\bm \delta}_\perp} 
          \int_{-\infty}^{\infty} dz\,
      \tr\Big[2{\bm \varepsilon}\cdot{\bm \gamma} 
         + \zeta\frac{d\bm \varepsilon}{d\zeta}
	\cdot{\bm \gamma}
-2{\bm \gamma}_0 
\Big]\Big|_{z^\prime=z}.
\end{equation}
It is evident that the second term in the energy expression (\ref{E-disp}) does not
vanish in this case. In fact, this term cancels the $\delta$-function piece coming from 
the first term. The self-energy for the TE mode is
\begin{equation}
({\cal E}-{\cal E}_0)^\mathrm{TE}=\frac{1}{8\pi^2}\int_0^\infty d\kappa \kappa^2
                            \int_{-1}^{1}d(\cos\beta)\cos^2\beta
\int_0^{2\pi} d\alpha \,e^{i\kappa{\delta}_{\perp} \sin\beta}\frac{\zeta_p}{\zeta_p+2\kappa},
\end{equation}
where $\kappa =\sqrt{k^2+\zeta^2}$. For an arbitrary coupling the TE energy per unit area 
is 
\begin{eqnarray}
({\cal E}-{\cal E}_0)^\mathrm{TE}&=& \frac{1}{8\pi^2\delta_\perp^3}\left\{\pi-
\sin\frac{\zeta_p\,\delta_\perp}{2}
\left[2\mbox{Ci}\left(\frac{\zeta_p\,\delta_\perp}{2}\right)
+\frac{\zeta_p\,\delta_\perp}{2}\left(\pi-2\mbox{Si}
\left(\frac{\zeta_p\,\delta_\perp}{2}\right)\right)\right]
\right . \nonumber\\
&& \hspace{16mm}\left . 
+\cos\frac{\zeta_p\,\delta_\perp}{2}\left[-\pi+2\frac{\zeta_p\,\delta_\perp}{2}
\mbox{Ci}\left(\frac{\zeta_p\,\delta_\perp}{2}\right)+2\mbox{Si}
\left(\frac{\zeta_p\,\delta_\perp}{2}\right)
\right]\right\},
\end{eqnarray}
where $\mbox{Ci}(x)$ and $\mbox{Si}(x)$ are the standard cosine integral and sine integral 
functions, 
respectively. For the finite coupling, $\zeta_p$, and in the $\delta_\perp \to 0$ limit we 
obtain the result identical to the divergence structure obtained in the analysis of the 
self-energy of a $\delta$-function plate interacting with a scalar 
field~\cite{Milton:2014-fall4}, 
\begin{equation}
({\cal E}-{\cal E}_0)^\mathrm{TE}_{\delta_\perp \to 0} =
\frac{\zeta_p}{8\pi^2\delta_\perp^2}\left\{1-\frac{\pi}{4}
\left(\frac{\zeta_p\,\delta_\perp}{2}\right)
+\frac{1}{3}\left(\frac{\zeta_p\,\delta_\perp}{2}\right)^2
\left[\frac{4}{3} -\gamma-\ln\left(\frac{\zeta_p\,\delta_\perp}{2}\right)\right]\right\}
+{\cal O}(\delta_\perp).
\label{TE-p-div}
\end{equation}
In the strong coupling limit $\zeta_p \to \infty$, keeping $\delta_\perp$ finite, we 
recover the inverse cubic divergence,
\begin{equation}
({\cal E}-{\cal E}_0)^\mathrm{TE}_{\lambda \to\infty}=\frac{1}{8\pi \delta_\perp^3}.
\label{div-st}
\end{equation}
The TM mode energy is
\begin{equation}
({\cal E}-{\cal E}_0)^\mathrm{TM}=\frac{1}{8\pi^2}\int_0^\infty d\kappa \kappa^2
                             \int_{-1}^{1}d(\cos\beta)(1+\sin^2\beta)
             J_0(\kappa \delta_\perp\sin\beta)\frac{\zeta_p}{\zeta_p+2\kappa\cos^2\beta}.
\end{equation}
In the strong coupling limit $\zeta_p \to\infty$, we obtain an inverse third power of 
point-splitting parameter as in (\ref{div-st}) with opposite sign. Thus in the strong 
coupling the total energy per unit area show no divergence.

To obtain the finite or weak coupling divergence structure for the TM, we first write the integrand as
\begin{equation}
({\cal E}-{\cal E}_0)^\mathrm{TM}=\frac{1}{8\pi^2}
\sum_{q=0}^{\infty}(-1)^q\left(\frac{\lambda}{2}\right)^{q+1}
\int_{-1}^{1}d(\cos\beta)\frac{(1+\sin^2\beta)}{(\cos\beta)^{(2q+2)}}
\int_0^\infty d\kappa \,\kappa^{(1-q)}                         
             J_0(\kappa \delta_\perp\sin\beta).
\end{equation}
We now carry out the $\kappa$ and $\cos\beta$ integrations for a fixed $k$,
\begin{equation}
({\cal E}-{\cal E}_0)^\mathrm{TM}=\frac{1}{8\pi^2}
\sum_{q=0}^{\infty}(-1)^q\left(\frac{\lambda}{2}\right)^{q+1}
\frac{1}{2^{(1+q)}\pi}q \,\delta^{-2+q}\cos\frac{q\pi}{2}\,\Gamma\left(-\frac12 - q\right)
\Gamma\left(-\frac q2\right)\Gamma\left(\frac q2\right),
\qquad \frac12<q<2 .
\end{equation}
We are only interested in looking at the divergence behavior of the TM self-energy of the 
$\delta$-function plate in the limit $\delta_\perp\to 0$, so we shall evaluate the above 
expression for $q=0,1,2$. The above expression has finite limit for $q=0,1$ and has a pole 
at $q=2$, which presumably could be removed by introducing a small photon mass. Keeping 
all the terms together we find,
\begin{equation}
({\cal E}-{\cal E}_0)^\mathrm{TM}_{\delta_\perp \to 0}=
\frac{\zeta_p}{8\pi^2\delta_\perp^2}\left\{1
+\frac{1}{15}\left(\frac{\zeta_p\,\delta_\perp}{2}\right)^2
\left[\frac{1}{q-2}+\ln\left(\frac{\zeta_p\,\delta_\perp}{2}\right)-\frac{\gamma}{2}-\ln 2
-\psi\left(-\frac52\right)+\frac12\psi\left(\frac32\right)\right]\right\}
+{\cal O}(\delta_\perp).
\label{TM-p-div}
\end{equation}
Notice that divergences in Eqs. (\ref{TE-p-div}) and (\ref{TM-p-div}) do not cancel 
between the TE and TM mode. We shall see a similar behavior for the finite coupling 
case of the $\delta$-function sphere.  
\subsection{Self energy of a electric $\delta$-function sphere}
\label{sp-self-el}

Next we consider the purely electric $\delta$-function sphere. With the choice of the 
plasma model described above, the TE and TM Green's functions obtained here coincide with 
those discussed in Refs.~\cite{Milton:2004vy,Milton:2004ya,Dalvit:2011edi,Milton:2002vm} with the 
identification of $\lambda^{\rm TE}=a^2\zeta_p=-x^2 \lambda^{\rm TM}$, and the 
redefinition of spherical Bessel function in terms of modified Riccati-Bessel function as 
$s_l=x\,{\text i}_l$ and $e_l=\frac2\pi x\,{\text k}_l$. Thus the results found there, 
with errors corrected in Ref.~\cite{Dalvit:2011edi}, for the energy and the stress on the 
sphere, follow with the above coupling constant identification.

The integral $I$ defined in Eq.~(\ref{I}) in this case becomes
\begin{equation}
I=\int_0^\infty r^2 dr \left[2{\bm \gamma}
                       +\frac{2}{\zeta^2}{\bm \zeta}_p\cdot{\bm \gamma}\delta(r-a)
                 -\frac{2}{\zeta^2} \zeta_p\cdot{\bm \gamma}\delta(r-a)
  -2{\bm \gamma}_0\right]
=\int_0^\infty r^2 dr 2\left({\bm \gamma}
                        -{\bm \gamma}_0\right),
\end{equation}
where the $\delta$-function terms coming from the first and the second terms cancel, 
similar to the $\delta$-function plate case. If we fail to take the dispersion term in 
account, then we would get additional contributions from the remaining $\delta$-function term. 


The integrals for the TE and TM mode are
\begin{subequations}
\begin{eqnarray*}
I^\mathrm{TE}&=&2\int_0^\infty r^2 dr\zeta^2\frac{2\zeta}{\pi}\left[\theta(a-r)
       \sigma^\text{scatt(E)}_{l,><}{\text i}^2_l(\zeta r)+
       \theta(r-a)\sigma^\text{scatt(E)}_{l,<>}{\text k}^2_l(\zeta r)\right],\\
I^\mathrm{TM}&=&2\int_0^\infty r^2 dr\frac{2\zeta}{\pi}\left[\theta(a-r)
       \sigma^\text{scatt(H)}_{l,><}\left(\zeta^2{\bar{\text i}}^2_l(\zeta r)
       +\frac{l(l+1)}{r^2}{\text i}^2_l(\zeta r)\right)
       +\theta(r-a)\sigma^\text{scatt(H)}_{l,<>}\left(\zeta^2{\bar{\text i}}^2_l(\zeta r)
       +\frac{l(l+1)}{r^2}{\text k}^2_l(\zeta r)\right)\right].
\end{eqnarray*}
\end{subequations}
The above expressions are valid for the general case where both electric and magnetic 
coupling can be present. From Eqs.~(\ref{scatt-coeff}), it is evident that the integrand 
$I$ will vanish identically. This implies that the self-energy of a $\delta$-function shell 
that is both perfectly electrically conducting and a perfectly magnetically conducting is 
zero including the divergences! Such a shell does not have any optical interaction up to a 
phase.  

Using the identities
\begin{subequations}
\begin{eqnarray}
\int_0^\infty dx\,x^2\, {\text i}_l^2(x)&=&\frac x2\left[(x^2+l(l+1)){\text i}^2
                             -x\,{\text i}_l{\text i}_l^\prime-x^2\,{\text i}_l^{\prime 2} \right],\\
\int_0^\infty dx\,x^2\, {\text k}_l^2(x)&=&-\frac x2\left[(x^2+l(l+1)){\text k}^2
                             -x\,{\text k}_l{\text k}_l^\prime-x^2\,{\text k}_l^{\prime 2} \right],
\label{stand-int}
\end{eqnarray}
\end{subequations}
and the Wronskian $W[{\text i}_l(x),{\text k}_l(x)]=-\frac{\pi}{2x^2}$, and keeping both 
temporal and spatial point-splitting we obtain the energy for the TE and TM mode as
\begin{subequations}
\begin{eqnarray}
E^\mathrm{TE}&=&
-\frac12\int_{-\infty}^\infty \frac{d\zeta}{2\pi}\frac{e^{i\zeta\tau}-1}
{i\zeta\tau}\sum_{l=1}^\infty (2l+1)P_l(\cos\delta)\zeta\frac{d}{d\zeta}\ln
        \left[1+\zeta_p a\frac2\pi x \,{\text i_l(x)}{\text k_l(x)}\right],
\label{te-sp}\\
E^\mathrm{TM}&=&-\frac12\int_{-\infty}^\infty \frac{d\zeta}{2\pi}\frac{e^{i\zeta\tau}-1}
{i\zeta\tau}\sum_{l=1}^\infty (2l+1)P_l(\cos\delta)\zeta\frac{d}{d\zeta}\ln
        \left[-1+\zeta_p a\frac2\pi x \,{\bar{\text i}_l(x)}_l{\bar{\text k}_l(x)}_l\right],
\label{tm-sp}
\end{eqnarray}
\end{subequations}
where $x=|\zeta| a$ for a sphere of radius $a$.

Thus the total self-energy of an electric $\delta$-function sphere is 
\begin{equation}
E=-\frac12\int_{-\infty}^\infty \frac{d\zeta}{2\pi}\frac{e^{i\zeta\tau}-1}
{i\zeta\tau}\sum_{l=1}^\infty (2l+1)P_l(\cos\delta)\zeta\frac{d}{d\zeta}
\ln\left[1+\zeta_pa\frac{e_l(x)s_l(x)}x\right]
\left[1-\zeta_pa\frac{e_l'(x)s_l'(x)}x\right],
\label{del-sph-tot-E}
\end{equation}
where we have used the prevalent modified Riccati-Bessel functions. 
In Ref.~\cite{Milton:2004ya,Dalvit:2011edi}, we tried to make sense of this expression 
without serious 
regulation.  Now, everything will be well-defined, and we shall study carefully the cutoff 
dependences. 


\subsubsection{Strong couping}

In the perfect conducting limit $\zeta_p\to\infty$ (strong coupling) we recover the 
standard result, which is the well-studied Boyer problem~\cite{Boyer:1968uf,Davies:1972css,Balian:1977qr,Milton:1978sf,Leseduarte:1996sac,Nesterenko:1998ces}.
\begin{eqnarray}
E-E_0&=&-\frac1{4\pi}\sum_{l=1}^\infty (2l+1)P_l(\cos\delta)\int_{-\infty}^\infty
d\zeta \frac{e^{i\zeta\tau}-1}{i\zeta\tau}\zeta\frac{d}{d\zeta}\ln\frac{
e_ls_le_l's_l'}{x^2}\\
&=&-\frac{1}{4\pi }\sum_{l=1}^\infty(2l+1)P_l(\cos\delta)
        \int_{-\infty}^\infty d\zeta\,\frac{e^{i\zeta\tau}-1}
{i\zeta\tau} \zeta 
        \left[\frac{s_l^\prime}{s_l}+\frac{e_l^\prime}{e_l}
          +\frac{s_l^{\prime\prime}}{s_l^\prime}
          +\frac{e_l^{\prime\prime}}{e_l^\prime}-\frac{2}{x}\right].\nonumber
\label{del-sph-PC}
\end{eqnarray}
Here, we carefully extract the divergent terms, and obtain the familiar finite remainder. 
First, we note that the $\ln (1/x)$ term does not contribute, because
\begin{equation}
\int_{-\infty}^\infty \frac{d\zeta}{i\zeta\tau}\left(e^{i\zeta\tau}-1\right)
=\frac2\tau\int_{-\infty}^\infty \frac{d\zeta}\zeta e^{i\zeta\tau/2}\sin\frac
{\zeta\tau}2=\frac\pi{\tau}, 
\end{equation}
which is a constant, independent of the size of the sphere, so the corresponding 
contribution to the energy is irrelevant.

To proceed, we use the uniform asymptotic expansions for the Bessel functions to find for 
large $\nu=l+1/2$,
\begin{equation}
\ln e_le_l's_ls_l'\sim -\ln 4-\frac{t^6}{4\nu^2}+\frac{t^6}{32\nu^4}
(4-54t^2+120t^4-71 t^6)+O(\nu^{-6}).
\end{equation}
Here $x=|\zeta|a=\nu z$ and $t=(1+z^2)^{-1/2}$. The order $\nu^{-2}$ term gives rise to a 
divergent contribution to the energy in the absence of a cutoff.  With the above cutoff, 
that term yields the energy contribution
\begin{equation}
E^{(2)}=\frac1{8\pi a}\sum_{l=1}^\infty P_l(\cos\delta)\int_{-\infty}^\infty
dz\frac{e^{i\nu z\tilde\tau}-1}{i\nu z\tilde\tau}z\frac{d}{dz}\frac1{(1+z^2)^3},
\end{equation}
with $\tilde \tau=\tau/a$. The $z$ integral is easily evaluated, leaving
\begin{eqnarray}
E^{(2)}&=&-\frac1{64a}\sum_{l=1}^\infty P_l(\cos\delta)e^{-\nu\tilde\tau}
(3+3\nu\tilde\tau+\nu^2\tilde\tau^2)\nonumber\\
&=&-\frac1{64a}\left(3-3\tilde\tau\frac{\partial}{\partial\tilde\tau}
+\tilde\tau^2\frac{\partial^2}{\partial\tilde\tau^2}\right)\sum_{l=1}^\infty
P_l(\cos\delta)e^{-\nu\tilde\tau}.
\end{eqnarray}
In the limit of small $\tau$ and $\delta$, the sum on $l$ is evaluated, using the 
generating function for the Legendre polynomials,
\begin{equation}
\sum_{l=1}^\infty P_l(\cos\delta)e^{-\nu\tilde\tau}=-1+\frac1\Delta,\quad
\Delta=\sqrt{\delta^2+\tilde\tau^2}.
\end{equation}
Thus the divergent term in strong coupling is
\begin{equation}
E^{(2)}=\frac{3}{64a}-\frac1{64a\Delta^5}(3\delta^4+8\delta^2\tilde\tau^2
+8\tilde\tau^4).
\label{E2-new-cutoff}
\end{equation}
Geometrically, the divergent term, as $\delta$ and $\tilde\tau$ tend to zero, corresponds 
to a surface integral of the curvature-squared divergence, which is uncanceled between the TE and TM modes, and 
between interior and exterior contributions.  On the other hand, the finite part, which 
arises entirely from the omitted $l=0$ term in the sum, $3/(64 a)$, is within 2\% of the 
exact repulsive result~\cite{Milton:1978sf}.

This result seems rather surprising, since the conventional wisdom is that this divergence 
is not present for a perfectly conducting spherical shell of zero 
thickness~\cite{Fulling:2003zx}. Indeed, the $a_2$ heat kernel coefficient for this problem 
vanishes.  To elucidate this conundrum, we note that the form of the temporal cutoff used 
here is a bit unconventional. What was actually used in the time-split regulated 
calculation in Ref.~\cite{Milton:1978sf} was simply $e^{i\nu z\tilde \tau}$ rather than 
$(e^{i\nu z\tilde\tau}-1)/i\nu z\tilde\tau$.  We can verify that if the above calculation  
is repeated for the former regulator, we instead find
\begin{equation}
E^{(2)\prime}
=\frac3{64 a}-\frac3{64a}\frac{\delta^2}{\Delta^7}(\delta^4+4\delta^2
\tilde\tau^2+8\tilde\tau^4).
\label{E2-old-cutoff}
\end{equation}
Now, if the spatial cutoff is set to zero, $\delta=0$, the divergent term vanishes! This 
seems to be the content of the heat-kernel approach.  And, in fact, with a purely spatial 
regulator,
\begin{equation}
E^{(2)}=\frac{3}{64 a}\left(1-\frac1\delta\right).\label{e2sp}
\end{equation}
This result was anticipated, for example, in Refs.~\cite{Milton:2013yqa,Milton:2013xia} 
(see also Refs.~\cite{Dowker:1995ffd,Apps:1998pbd,Nesterenko:2003hkc}) where it was found 
for a single curvature, the TE and TM  integrated curvature squared divergent 
contributions are for an arc of angle $\alpha$ 
\begin{equation} 
\mbox{TE}:\quad -\frac{1}{\pi}\frac{\alpha}{1024 a\delta},\quad
 \mbox{TM}:\quad -\frac{1}{\pi}\frac{5\alpha}{1024 a\delta},
\end{equation}
so when $\alpha=2\pi$, the sum of these two multiplied by 4 (two curvatures, and inside 
and outside contributions) yields the divergence found in Eq.~(\ref{e2sp}), and further 
the ratio of the TE and TM contributions, 1/5, is indeed found here when the individual contributions are examined. (See Eq.~(\ref{ques}), below.)

Now to extract the finite part, we can follow the procedure given in 
Refs.~\cite{Bender:1994zr,Milton:1996ri}, and use the following asymptotic evaluations of 
the integrals, with $\tau=0$,
\begin{subequations}
\label{ques}
\begin{eqnarray}
Q_l^{\rm TM}&=&\int_0^\infty dx \ln(-2e_l's_l')\sim\frac{\pi\nu}2
-\frac{5\pi}{128\nu}-\frac{53\pi}{32769\nu^3}-\frac{901\pi}{2097152\nu^5}+\dots,
\\
Q_l^{\rm TE}&=&\int_0^\infty dx \ln(2e_ls_l)\sim-\frac{\pi\nu}2
-\frac{\pi}{128\nu}+\frac{35\pi}{32769\nu^3}-\frac{565\pi}{1048576\nu^5}+\dots,
\end{eqnarray}
\end{subequations}
so the total integral here is
\begin{equation}
Q_l=Q_l^{\rm TE}+Q_l^{\rm TM}\sim-\frac{3\pi}{64\nu}-\frac{9\pi}{16348
\nu^3}-\frac{2031\pi}{2097152\nu^5}+\dots.\label{qas}
\end{equation}
Thus with spatial regulation the energy has the form
\begin{equation}
E=\frac1{2\pi a}\sum_{l=0}^\infty (2l+1) P_l(\cos\delta)Q_l+\frac\pi{48a},
\label{pcsp}
\end{equation}
where it is convenient to start the sum at $l=0$, so we subtract off the value of that 
term.  Putting in the first three terms of the asymptotic expansion gives us
\begin{equation}
E-\frac\pi{48a}\sim -\frac{3}{64 a\delta}-\frac1a\left(\frac{9\pi^2}{32768}
+\frac{677\pi^4}{4194304}\right)=-\frac{3}{64a\delta}-\frac{0.0184335}{a}.
\end{equation}
To this we must add the remainder, obtained from Eq.~(\ref{pcsp}) by subtracting the first 
three asymptotic terms given in (\ref{qas}) from $Q_l$: 
\begin{equation}
R=\frac1{\pi a}\sum_{l=0}^\infty \nu\left(Q_l+\frac{3\pi}{64\nu}+\frac{9\pi}
{16384\nu^3}+\frac{2031\pi}{2097152\nu^5}\right).
\end{equation}
The sum converges rapidly; going out to $l=2$ is sufficient to give $R=-0.000840$, giving 
us for the energy
\begin{equation}
E=-\frac3{64 a\delta}+\frac{0.046176}a,
\end{equation}
where the finite part is the standard number for a perfectly conducting sphere. Note that 
the first approximation we had in Eq.~(\ref{E2-new-cutoff}) is high by only 1.5\%.


\subsubsection{Finite Coupling}\label{sec:fc}
We now return to the general expression (\ref{del-sph-tot-E}), and start by analyzing the 
divergences occurring there as the cutoff parameters $\delta$ and $\tau$ approach zero. In 
doing so, we again use the uniform asymptotic expansion for the Bessel functions; we 
immediately encounter a difficulty in that the TM mode contributions yield a spurious 
infrared divergence, because of the $1/z$ behavior for small $z$.  To cure this, we insert 
an infrared cutoff as well: We replace 
\begin{equation}
\frac1{z^2 t}=\frac{\sqrt{1+z^2}}{z^2}\to \frac{\sqrt{z^2+1}}{z^2+1}=t;
\end{equation}
we could insert an arbitrary infrared cutoff parameter, but that introduces unnecessary 
complications, since in principle we will be adding back the same terms that we subtract. 
This way we can treat the TE and TM modes on the same footing.

The divergences occur in the asymptotic expansion of the logarithm in 
Eq.~(\ref{del-sph-tot-E}), which we write as
\begin{equation}
\ln\left(1+\frac{\zeta_pa}{x}e_ls_l\right)\left(1-\frac{\zeta_pa}{x}
e_l's_l'\right)\sim\sum_{k=1}^\infty \frac{a^{(k)}}{(2\nu)^k},\label{as}
\end{equation}
where a simple calculation gives
\begin{subequations}\label{valueas}
\begin{eqnarray}
a^{(1)}&=&2\zeta_p a t,\\
a^{(2)}&=&-\zeta_p^2a^2t^2,\\
a^{(3)}&=&\frac{\zeta_pa}3(-3t^7+2\zeta_p^2a^2 t^3),\\
a^{(4)}&=&\frac{\zeta_p^2a^2}2(2t^8-\zeta_p^2a^2t^4),
\end{eqnarray}
\end{subequations}
which result from remarkable cancellations between the individual mode contributions. Let 
us label the contributions from each term in the asymptotic series by $E^{(k)}$.  Since 
the integrals are regulated, we may integrate by parts, to obtain
\begin{equation}
E^{(k)}=\frac1{8\pi a}\sum_{l=1}^\infty (2l+1)^{2-k}P_l(\cos\delta)
\int_{-\infty}^\infty dz\, e^{i\nu z\tilde\tau}a^{(k)}(z).
\end{equation}
Then the first divergent term is
\begin{equation}
E^{(1)}=\frac{\zeta_p}{4\pi}\int_{-\infty}^\infty \frac{dz}{\sqrt{1+z^2}}2\frac
{d}{diz\tilde \tau}\sum_{l=1}^\infty P_l(\cos\delta)e^{i\nu z\tilde\tau},
\label{e1}
\end{equation}
where the $l$ sum evaluates to 
\begin{equation}
g(z\tilde\tau,\delta)=\sum_{l=1}^\infty P_l(\cos\delta)e^{i\nu z\tilde\tau}=
-e^{iz\tilde\tau/2}+\frac1{\sqrt{2(\cos z\tilde\tau-\cos\delta)}},\label{gee}
\end{equation}
where the first term comes from the omitted $l=0$ term. The corresponding energy can be 
evaluated to 
\begin{equation}
E^{(1)}_1=-\frac{\zeta_p}{4\pi}\int_{-\infty}^\infty dz\frac{\cos z\tilde \tau/2}
{\sqrt{1+z^2}}=-\frac{\zeta_p}{2\pi}K_0(|\tilde\tau|/2)\sim 
\frac{\zeta_p}{\pi}\left(\frac12\ln\frac{\tilde\tau}{4}+\frac{\gamma}{2}\right).\label{eone}
\end{equation}
To obtain the divergent contribution from the second part in Eq.~(\ref{e1}), we first set the 
spatial cutoff $\delta=0$ and integrate by parts to get
\begin{equation}
E^{(1)}_2=\frac{\zeta_p}{4\pi \tilde{\tau}}\int_{-\infty}^\infty dz\frac{z}{(1+z^2)^{3/2}}
\frac{1}{\sin z\tilde{\tau}/2}.
\label{e(1)-2}
\end{equation}
In the $\tilde\tau \to 0$ limit, if we keep only the first term in the expansion of the 
sine, we get
\begin{equation}
E^{(1)}_{2a}=\frac{\zeta_p}{\pi}\frac{1}{\tilde\tau^2},
\label{e(1)-2a}
\end{equation}
which the expected quadratic divergence. However, Eq.~(\ref{e(1)-2}) is not well defined, 
because it possesses infinite numbers of poles along the real axis. The proper interpretation 
is that the integral be understood as the principal part from each pole. (The poles, 
however encircled, would give an imaginary part.) We can write the pole structure for 
$z\tilde\tau > 0$ as 
\begin{equation}
\frac{1}{\sin z\tilde{\tau}/2}-\frac{1}{ z\tilde{\tau}/2}+\frac{1}{ z\tilde{\tau}/2}
=\frac{1}{ z\tilde{\tau}/2}+\sum_{n=1}^\infty\frac{(-1)^n}{z\tilde{\tau}/2 - n \pi} +
f(z\tilde{\tau}/2),
\end{equation}
where $f(z\tilde{\tau}/2)$ has no singularities. The first term gives the contribution 
given in Eq.~(\ref{e(1)-2a}). After carrying out the principal part integral of the pole 
at $z=2n\pi/\tilde{\tau}$, and then carrying out the sum on $n$ we obtain
\begin{equation}
E^{(1)}_{2b}=\frac{\zeta_p}{\pi}\left[ \frac{1}{2\pi \tilde{\tau}}\ln 2 
-\frac{1}{48}-\frac{1}{48}\ln\frac{\tilde{\tau}}{4\pi}-\frac{1}{48}\ln 2 -
\frac{1}{8\pi^2}\zeta^\prime(2)\right].
\label{e(1)-2b}
\end{equation}
To evaluate the contribution from the remainder function $f(z\tilde{\tau}/2)$, we first 
note that for small argument, $f(x) \sim -\frac{1}{\pi}\ln 2+\frac{x}{12}$, where the 
contribution from the first part will exactly cancel the inverse $1/\tilde{\tau}$ 
divergence in (\ref{e(1)-2b}). The contribution from the second part can be obtained by 
splitting the integral at L, where $1 \gg L \gg \tilde{\tau}/2$, and the integration from 
$L$ to $\infty$ needs to be verified numerically. Combining all pieces together we get 
the contribution from the remainder term as
\begin{equation}
E^{(1)}_{2c}=\frac{\zeta_p}{\pi}\left[ -\frac{1}{2\pi \tilde\tau}\ln 2-\frac{1}{48}
-\frac{1}{48}\ln\frac{\tilde\tau}{4}+\frac{0.08513}{4}\right].
\label{e(1)-2c}
\end{equation}
Adding all the contribution from Eqs.~(\ref{eone}), (\ref{e(1)-2a}),~(\ref{e(1)-2b}), 
and~(\ref{e(1)-2c}), we obtain the first order asymptotic term as
\begin{equation}
E^{(1)}=\frac{\zeta_p}{\pi}\left[\frac{1}{\tilde\tau^2}+\frac{11}{24}\ln\tilde\tau
-0.345879 \right].
\end{equation}

It is clear from (\ref{eone}) that the spatial cutoff alone will not render the integral 
convergent.  The $E^{(1)}$ term is 
usually omitted as being merely a ``tadpole'' term, in the language of perturbative (in 
$\zeta_p$) Feynman diagrams.

The second order term can be evaluated by either doing the $z$ integral or the $l$ sum 
first. In the former case,
\begin{eqnarray}
E^{(2)}&=&
-\frac{\zeta_p^2a}{8\pi}\sum_{l=1}^\infty P_l(\cos\delta)\int_{-\infty}^\infty 
dz \, e^{i\nu z\tilde\tau}t^2\nonumber\\
&=&-\frac{\zeta^2a}{8}\sum_{l=1}^\infty P_l(\cos\delta)e^{-\nu\tilde\tau}
=-\frac{\zeta_p^2a}{8}\left(-1+\frac1\Delta\right),\label{etwo}
\end{eqnarray}
where the last holds for very small $\tau$ and $\delta$, and 
$\Delta=\sqrt{\tilde\tau^2+\delta^2}$.
Like the similar divergent term that appeared in strong coupling, 
Eq.~(\ref{E2-new-cutoff}), this is a surface-integrated curvature term.

Once again, the appearance of this divergence in $E^{(2)}$ may cause surprise. It is not 
apparent in the heat kernel analysis~\cite{Bordag:1998vs}.  It was also not found in 
earlier analyses for the finite coupling scalar problem for the sphere~\cite{Milton:2004vy,Milton:2002vm}, 
which disagreed with calculations by Graham et al.~\cite{Graham:2002fw,Graham:2002xq}. But 
implicit in the earlier null results was, like we saw in strong coupling, a conventional 
time-splitting regulator. Indeed, if we repeat the above calculation with only a 
conventional exponential point splitting, we find
\begin{eqnarray}
E^{(2)}&=&-\frac1{8\pi a}\sum_{l=1}^\infty \int_{-\infty}^\infty
dz\,e^{i\nu z\tilde \tau}z\frac{d}{dz}(-\zeta_p^2a^2t^2)\nonumber\\
&=&\frac{\zeta^2_p a}8\sum_{l=1}^\infty (\nu|\tilde\tau|-1)e^{-\nu\tilde\tau}
=\frac{\zeta_p^2 a}8\frac{d}{d\tau}\tau
\left( e^{-\tilde\tau/2}-\frac1{2\sinh\tilde
\tau/2}\right)=\frac{\zeta_p^2 a}8,
\end{eqnarray}
as $\tilde\tau\to0$.  That is, we recover precisely the same finite part seen in 
Eq.~(\ref{etwo}), but not the divergent term!  As in strong coupling, conventional temporal 
point-splitting hides the divergence here. 

The last divergent contribution comes from $a^{(3)}$:
\begin{equation}
E^{(3)}=\frac1{8\pi}\sum_{l=1}^\infty
\frac{P_l(\cos\delta)}{2l+1}\int_{-\infty}
^\infty dz e^{i\nu z\tilde \tau}\frac{\zeta_p}3
\left[-3t^7+2\zeta_p^2a^2 t^3\right].
\end{equation}
The sum on $l$,
\begin{equation}
f(z\tilde\tau,\delta)=\sum_{l+1}^\infty \frac{P_l(\cos\delta)}{2l+1}e^{i(2l+1)
z\tilde\tau/2},
\end{equation}
may be evaluated by integrating Eq.~(\ref{gee}),
\begin{equation}
2\frac\partial{\partial i z\tilde\tau} f(z\tilde\tau,\delta)=g(z\tilde\tau,
\delta).
\end{equation}
For small $\tilde\tau$ amd $\delta$ there are two branches:
\begin{subequations}
\begin{eqnarray}
f(z\tilde\tau,\delta)&\sim&\frac32\ln 2-1-\frac12\ln\delta
+\frac{i}2\arcsin\frac{z\tilde\tau}\delta,\quad \delta>|z\tilde\tau|,\\
&\sim&\frac32\ln 2-1-\frac12\ln\delta+\mbox{sgn}(z)\left[\frac{i\pi}{4}
+\frac12\mbox{arccosh}\frac{|z\tilde\tau|}{\delta}\right],
\quad \delta<|z\tilde\tau|.
\label{e3-branches}
\end{eqnarray}
\end{subequations}
For the purely spatial cutoff, i.e. $\tilde\tau=0$, we obtain
\begin{equation}
E^{(3)}\sim \frac{\zeta_p}{6\pi}\left(-\frac45+\zeta_p^2a^2\right)
\left(\frac32\ln2-1-\frac12\ln\delta\right),
\label{E(3)}
\end{equation}
which is identical to the result obtained if we carry out the sum and integral in the 
opposite order~\cite{Milton:2017sp-ent}. Notice from Eq.~(\ref{e3-branches}) that we cannot 
set $\delta=0$ in this case because of the appearance of the $\ln \delta$ term. Apart from 
the different cutoff, however, the order $\zeta_p$ term rescales the logarithmic divergence 
already seen in $E^{(1)}$, Eq.~(\ref{eone}).  Again, with the same caveat, the 
$O(\zeta_p^3)$ term is that seen 
previously~\cite{Milton:2004vy,Milton:2004ya,Milton:2002vm}, corresponding to the familiar nonzero $a_2$ 
heat kernel coefficient found in Ref.~\cite{Bordag:1998vs}.

The occurrence of this logarithmic divergence, of course, was to be expected. It would 
seem to pose a barrier to computing a finite Casimir energy for an 
electromagnetic $\delta$-function sphere, because one could multiply the cutoff  $\delta$ 
by an arbitrary number, which would then change the finite part.  Only the strong-coupling 
limit, which is essentially that of $\varepsilon\to\infty$, yields a computable energy, 
that might, somehow, have observable consequences.

We notice the similar logarithmic divergence 
occurring in the finite coupling case of the $\delta$-function plate self-energy, shown in 
Eqs.~(\ref{TE-p-div}) and (\ref{TM-p-div}). In the asymptotic flat plate limit, the 
self-energy of a spherical shell leads to the self-energy of a $\delta$-function plate. 

\section{Stress on $\delta$-Sphere}
\label{sec4}

The electromagnetic stress tensor is
\begin{equation}
T^{\mu\nu}=F^{\mu\lambda}F^\nu{}_\lambda-\frac14 g^{\mu\nu}F^{\alpha\beta}
F_{\alpha\beta},
\end{equation}
so, in particular, the radial-radial component of the stress tensor is
\begin{equation}
T_{rr}=-\frac12 E_r^2-\frac12B_r^2+\frac12 E_\perp^2+\frac12 B_\perp^2.
\end{equation}
The Green's dyadic construction (\ref{dyadic-VSH}) leads to a TE and TM decomposition of 
the pressure on the spherical surface.  All that is needed to work out the components is 
the identity
\begin{equation}
\nabla_\perp^2 Y_{lm}(\theta,\phi)=-\frac{l(l+1)}{r^2}Y_{lm}(\theta,\phi).
\end{equation}
Then we obtain results nearly the same as those found earlier in 
Refs.~\cite{Milton:2004vy,Milton:2004ya,Dalvit:2011edi,Milton:2002vm}. The difference arises in  the TE 
modes which is, without the regulators inserted,
\begin{equation}
T_{rr}^{\rm TE}=\frac1{8\pi}\int_{-\infty}^\infty\frac{d\zeta}{2\pi}\sum_l
(2l+1)\left[-\zeta^2-\frac{l(l+1)}{r^2}+\frac1r\frac{\partial}{\partial r} r\frac1{r^\prime}
\frac{\partial}{\partial r^\prime}r^\prime\right]g^E_l(r,r^\prime)\bigg|_{r^\prime=r}.\label{stressform}
\end{equation}
The scalar Dirichlet case replaced the latter derivatives by
\begin{equation}
\frac1r\frac{\partial}{\partial r} r\to\frac{\partial}{\partial r}.
\end{equation}
Only the derivative terms contribute to the discontinuity in the stress, so we are left 
with the total outward stress on the sphere
\begin{equation}
S^{\rm TE}=4\pi a^2 T_{rr}\bigg|_{r=r'=a-}^{r=r'=a+}=
-\frac{\zeta_p}{2\pi a}\sum_{l=1}^\infty (2l+1)\int_{0}^\infty
dx\frac{(e_ls_l)'}{1+\zeta_p a\frac{e_ls_l}x}.
\end{equation}
The reason for the discrepancy with the earlier-derived 
result~\cite{Milton:2004vy,Dalvit:2011edi} is connected with the fact 
that this result is obtained from the unregulated form of the TE part of the energy 
(\ref{del-sph-tot-E}) when differentiated with respect to $a$, holding $\zeta_p$ fixed. 
(The form in Ref.~\cite{Dalvit:2011edi}  was obtained for $\lambda^{\rm TE}=\zeta_p a^2$ 
held fixed.)

How does this work in the presence of the regulators, where the TE energy is
\begin{equation}
E^{\rm TE}=-\frac1{4\pi}\int_{-\infty}^\infty d\zeta
\frac{e^{i\zeta\tau}-1}{i\zeta
\tau}\sum_{l=1}^\infty (2l+1)P_l(\cos\delta)\zeta\frac{d}{d\zeta}\ln
\left(1+\zeta_p a\frac{e_ls_l}x\right)?
\end{equation}
If we differentiate with respect to $-a$, and integrate by parts on $\zeta$, we 
immediately obtain
\begin{equation}
-\frac{\partial E^{\rm TE}}{\partial a}=
-\frac{\zeta_p}{4\pi a}\int_{-\infty}^\infty dy e^{iy\tilde\tau}
\sum_{l=1}^\infty(2l+1)P_l(\cos\delta)\frac{(e_ls_l)'}
{1+\zeta_p a \frac{e_ls_l}{x}}=S^{\rm TE}.
\end{equation}
Here, after differentiation, we have changed the integration variable to $y=\zeta a$, so 
that $\zeta\tau=y\tau/a=y\tilde \tau$, and as before, $x=|y|$. This is exactly the 
regulation we expect for the stress, which originates from the vacuum expectation value of 
the radial-radial component of the stress tensor. Thus the use of the elaborated temporal 
regulator seems vindicated.

This also works for the TM stress, which now exactly coincides with that found 
earlier~\cite{Milton:2004vy,Dalvit:2011edi}. 
The regulated form of the TM energy is
\begin{equation}
E^{\rm TM}=-\frac1{4\pi}\int_{-\infty}^\infty d\zeta \frac{e^{i\zeta\tau}-1}{i
\zeta\tau}\sum_{l=1}^\infty (2l+1)P_l(\cos\delta)\zeta\frac{d}{d\zeta}\ln
\left(1-\frac{\zeta_p}\zeta e_l's_l'\right),
\end{equation}
so again when this is differentiated with respect to $-a$, and integrated by parts in 
$\zeta$, we obtain the expected regulated TM stress:
\begin{equation}
-\frac{\partial E^{\rm TM}}{\partial a}=\frac{\zeta_p}{4\pi a}
\int_{-\infty}^\infty dy\, e^{iy\tilde \tau} \sum_{l=1}^\infty (2l+1)
P_l(\cos\delta)\frac{(e_l's_l')'}
{1-\frac{\zeta_p a}{x} e_l's_l'}=S^{\rm TM}.
\end{equation}
The latter may also be obtained from Eq.~(\ref{stressform}) with the replacement 
$g^E\to g^H$.


\section{Electric and Magnetic Couplings}
\label{sec5}

Finally we turn to the examination of the situation when both electric and magnetic 
couplings are present.  According to the results of Sec.~\ref{delta-sphere} the energy in 
general is given by
\begin{equation}
E-E_0=-\frac12\sum_{l=1}^\infty (2l+1)P_l(\cos\delta)\int_{-\infty}^\infty
\frac{d\zeta}{2\pi}\frac{e^{i\zeta\tau}-1}{i\zeta\tau}\zeta\frac{d}{d\zeta}
\ln\Delta^E\Delta^H,\label{generalends}
\end{equation}
where (the $\perp$ superscript on the couplings is omitted here)
\begin{subequations}
\begin{eqnarray}
\Delta^E&=&1+\frac{\zeta^2}4\lambda_e\lambda_g+|\zeta|(\lambda_e e_ls_l-\lambda_g
e'_ls'_l),\\
\Delta^H&=&1+\frac{\zeta^2}4\lambda_e\lambda_g+|\zeta|(\lambda_g e_ls_l-\lambda_e
e'_ls'_l),
\end{eqnarray}
\end{subequations} 
which generalizes Eq.~(\ref{del-sph-tot-E}) and has the expected symmetry between the 
electric and magnetic couplings.  In deriving this result, the following dispersion 
relations were assumed, a generalized plasma model,
\begin{equation}
\lambda_e=\frac{\zeta_p}{\zeta^2},\quad \lambda_g=\frac{\zeta_m}{\zeta^2}.
\label{plasma-coupling}
\end{equation}
From this form it is apparent that a purely magnetically coupled $\delta$-sphere behaves 
precisely the same as a purely electrical one, just the E and H modes are interchanged.

Let us now use the uniform asymptotic expansion to extract the leading behavior of the 
logarithm in the energy (\ref{generalends}).  The calculation follows very closely that 
summarized in Sec.~\ref{sec:fc}.  In particular, since we are interested only in 
asymptotic behavior, as there, we replace $z\to1/t$ asymptotically in the $e_l's_l'$ 
terms. Then using the notation of Eq.~(\ref{as}), 
\begin{equation}
\ln\Delta^E\Delta^H\sim\sum_{l=1}^\infty \frac{a^{(k)}}{(2\nu)^k},
\end{equation}
we obtain, with the abbreviations $\lambda=\zeta_p a$, $\hat\lambda=\zeta_m a$,
\begin{subequations}
\begin{eqnarray}
a^{(1)}&=&2t(\lambda+\hat\lambda),\\
a^{(2)}&=&-t^2(\lambda^2+\hat\lambda^2),\\
a^{(3)}&=&-t^7(\lambda+\hat\lambda)+\frac23 t^3(\lambda^3+\hat\lambda^3),\\
a^{(4)}&=&t^8(\lambda+\hat\lambda)^2-\frac12t^4(\lambda^4+\hat\lambda^4).
\end{eqnarray}
\end{subequations}
This generalizes Eqs.~(\ref{valueas}).
It is interesting that the first interference term between the electric and magnetic term 
occurs in the fourth coefficient, which means that the terms corresponding to the 
divergences exhibit no such interference.  This further means that it is impossible to 
find values of the couplings that will allow us to extract a finite energy.  That is, the 
divergences found in Sec.~\ref{sec:fc} are the same here, but with obvious changes in the 
coupling constant dependence.  There is one exception to this statement of impossibility, 
when
\begin{equation}
\lambda=-\hat\lambda.
\label{isorefractive}
\end{equation}
Then the first and last divergent terms vanish, and it would be possible to isolate a 
unique finite part.  This corresponds to the familiar cancellation that occurs for a 
dielectric-diamagnetic ball with the same speed of light inside and outside, that is 
$\varepsilon\mu=1$~\cite{Brevik:1983pt}.  For further details on this scenario see 
Ref.~\cite{Milton:2001zpe}.


\section{Conclusions}
\label{concl}

In this paper, we have extended the electromagnetic $\delta$-function potential 
formalism~\cite{Parashar:2012it,Milton:2013bm} to spherical geometry.  We modeled the 
spherical shell with the electric susceptibility $\bm \varepsilon - \bm 1$ and the magnetic 
susceptibility $\bm \mu - \bm 1$ using a $\delta$-function. We have unambiguously obtained 
the boundary conditions by integrating Maxwell's equations across the $\delta$-function 
sphere. Like in the case of the $\delta$-function plate, the polarizability 
$\lambda^{||}$, corresponding to the radial polarizabilities in the spherical case,  
are forbidden. We have provided further argument in favor of this 
observation. The perfect conductor limit is achieved by taking the limit 
$\lambda_e \to \infty$ for the purely electric $\delta$-function sphere, which corresponds 
to the perfectly conducting spherical shell as considered by Boyer~\cite{Boyer:1968uf}. In 
a similar limit, where both electric coupling  $\lambda_e \to \infty$ and magnetic coupling 
$\lambda_g \to \infty $ is taken, the the self-energy of the perfectly conducting 
magneto-electric $\delta$-function shell identically vanishes. The spherical shell in this 
case is transparent with the transmission coefficient showing a phase change $\pi$.

The finite coupling takes dispersion into account. The necessity of specifying the 
frequency dependence of the couplings, representing the permittivity and the permeability 
of the shell, is a consequence of the general formalism employed, which requires knowledge 
of the dispersion. In this paper, we have used a plasma like model.

When there is only an electric coupling, the formulas obtained 
earlier~\cite{Milton:2004vy,Milton:2004ya,Dalvit:2011edi,Milton:2002vm} are reproduced, but now with a 
definite relation between the TE and TM coupling constants.  In the present 
work we examine the divergence structure carefully.  We first do so in the strong 
coupling limit, where we reproduce the classic Boyer result~\cite{Boyer:1968uf,Milton:1978sf}, 
but now with a curvature-squared divergent term, which accidentally cancels when only a 
simple exponential time-splitting regulator is used, the latter corresponding to the familiar 
heat-kernel result.  For finite coupling, we compute the first three leading contributions 
resulting from the uniform asymptotic expansion of the modified Bessel functions; all 
three give divergent contributions.  The first, $O(\nu^{-1})$, contribution, diverges as 
the logarithm of the temporal point-splitting parameter; this term may regarded as a 
constant, and disregarded as a ``tadpole'' contribution.  The second, $O(\nu^{-2})$, 
contribution, leading to an inverse-linear dependence on the cutoff parameters, is 
expected as a curvature-squared divergence once again, but at least can be uniquely 
isolated. (For apparently accidental reasons this divergence again cancels for the simple 
exponential point-split temporal regulator, which is why it does not show up in heat 
kernel analyses.) But in the third term, of $O(\nu^{-3})$, a logarithmic divergence occurs 
which can only be regulated by a spatial point-split regulator.  Because of the scale 
ambiguity of such a logarithmic term, it is impossible to subtract it off, and therefore 
impossible, apparently, to compute a finite remainder.

This is in contrast to some earlier papers that obtain seemingly discordant results. Graham, 
Quandt, and Weigel~\cite{Graham:2013yza} consider a dielectric shell, characterized by a 
Drude-type dispersion relation, and a profile function.  Although they are unable to find 
a finite energy for such a shell, they can tune the profile function with the radius of 
the sphere so that the difference in energies between two such spherical bodies is finite, 
and thereby compute a unique force.  This procedure seems artificial, and further they do 
not correctly incorporate dispersion in their formalism as we do here. In any case, since 
our shell is a $\delta$ function, we have no profile to tune.

Another, even more recent paper, is by Beauregard, Bordag, and 
Kirsten~\cite{Beauregard:2015joa}.  They consider a $\delta$-function potential, and claim the 
divergences can be uniquely subtracted, in contradistinction to statements by two of the 
same authors, using a similar analysis, many years ago~\cite{Bordag:1998vs}. The new 
argument, based on the same heat kernel expansion given earlier, is that the divergent 
terms depend on positive powers of the mass, so must be ``renormalized away'' by the 
requirement that the Casimir energy must vanish as the mass of the field goes to infinity. 
This is not consistent with the conventional understanding of renormalization.  It also 
does not seem possible to adapt this idea here, since we deal with electromagnetism from 
the outset, which must be characterized by a massless photon field.

So whatever the merits of these new proposals, they are without bearing on our problem. We 
have encountered a difficulty in extracting a finite Casimir energy for a sphere for a 
purely electric $\delta$-function except in the special case of a perfectly conducting 
shell, what we are terming strong coupling. There are fascinating features noted in the 
cases of the spherical shell having both electric and magnetic properties; both in the 
perfectly conducting case, as mentioned above, and the finite coupling case. In the 
exceptional case, when the electric and magnetic couplings are equal and opposite, the 
contributions of the odd orders in coupling in the asymptotic expansions vanish, which 
mimics the case of the perfect conductor where a finite result can be obtained. We shall 
discuss this case elsewhere.

 
\acknowledgments
This work reported here  was supported in part by grants from the Julian Schwinger
Foundation and the Simons Foundation.  Some of the work was carried out at Laboratoire 
Kastler Brossel, CNRS, ENS, UPMC, Paris, which we further thank for hospitality and 
support. We acknowledge the support from the Research Council of Norway (Project No. 
250346). We thank Jef Wagner for assistance in some of the numerical calculations and 
Steve Fulling for helpful comments.


\appendix
 
\section{On a possible connection between cutoff parameter and  surface pressure}

There is reason to believe that among the various cutoff dependent terms it is the second 
order contribution that is of physical significance. The first order terms are usually 
considered to be without physical meaning, and when it comes to higher order terms, 
logarithms of cutoff parameters seem to be beyond measurability, even in principle. 
Quantities lacking a possibility of experimental test should naturally  be  deemed to be 
mathematical artifacts, simply reflecting the crudeness of the underlying physical model.
The delta sphere model in our case may be considered to be an example of that sort. When 
it comes to the second order energy  $E^{(2)}$, however, it is easy to see that it is much 
closer  to physical reality as it is from dimensional reasons closely connected with the 
concept of a {\it surface pressure}, obviously a concept having  physical meaning.

We shall now elaborate on this idea in more detail, working from here with dimensional 
units.

\subsection{ Strong coupling}

With use of the ``new'' cutoff parameter $(\exp{(i\nu z\tilde{\tau}})-1)/i\nu z\tilde{\tau}$, 
and setting $\delta=0$, one sees that Eq.~(\ref{E2-new-cutoff}) reduces to
\begin{equation}
E^{(2)}=\frac{3\hbar c}{64a}\left( 1-\frac{8}{3\tilde{\tau}}\right). \label{A1}
\end{equation}
With instead using  the traditional cutoff parameter $\exp(i\nu z\tilde{\tau})$ one has from Eq.~(\ref{E2-old-cutoff}), assuming a purely spatial cutoff ($\tau=0$),
\begin{equation}
E^{(2)}=\frac{3\hbar c}{64a}\left( 1-\frac{1}{\delta}\right). \label{A2}
\end{equation}
The important terms in the present context are the cutoff terms, which are seen to be large and negative, corresponding to an inward force. The above two expressions would simply be equivalent, if   $\tilde{\tau}=\tau c/a$   could be assumed to be a constant.  Such an assumption would not comply with our treatment above, however, which assumed that $\tau$, not $\tilde{\tau}$, is constant. We therefore choose to start from Eq.~(\ref{A1}), replacing $E^{(2)}$ by its cutoff dependent part. Thus,
\begin{equation}
E^{(2)} \rightarrow -\frac{\hbar c}{8}\frac{1}{\tau}. \label{A3}
\end{equation}
This term is a constant, giving zero when differentiated with respect to $a$. That is, this case does not correspond to a surface tension at all.

 We  move on to the case of finite coupling which, at least at first sight,  should be a more  natural situation in relation to  the surface pressure concept.


\subsection{Finite coupling}

We focus again on the cutoff dependent part in Eq.~(\ref{etwo}),
\begin{equation}
E^{(2)}=-\frac{\zeta_p^2 a\hbar}{8c}\frac{1}{\Delta}, \label{A7}
\end{equation}
and include, as above, only the temporal cutoff so that $\Delta \rightarrow \tau c/a$.

Differentiating with respect to $a$  keeping $\tau$ constant, we calculate 
\begin{equation}
\frac{\partial E^{(2)}}{\partial a}=-\frac{\zeta_p^2a\hbar}{4c}\frac{1}{c\tau}. 
\end{equation}
It corresponds to the surface pressure
\begin{equation}
f= -\frac{1}{4\pi a^2}\frac{\partial E^{(2)}}{\partial a}= +\frac{\zeta_p^2\hbar}{16\pi ac}\frac{1}{c\tau}. \label{A8}
\end{equation}
Remarkably enough,  this force acts {\it outwards}. The derivative with respect to $a$ does not in this case change the sign of the expression.

If we nevertheless proceed to equate $f$ to the hydrodynamical surface pressure $4\sigma/a$ for a fluid shell ("soap-bubble" geometry), we obtain
\begin{equation}
\sigma=-\frac{\zeta_p^2\hbar}{64 \pi c}\frac{1}{c\tau}. \label{A9}
\end{equation}
  Remarkably enough, we  see that  $\sigma \propto 1/\tau$, independently of the value of 
  $a$.  Thus there is an analogy to the result recently given in Ref.~\cite{Hoye:2017rst}, 
dealing with the surface pressure on a dielectric fluid ball. That derivation was based 
upon the earlier quantum field theory given in Ref.~\cite{Milton:1979yx} for the Casimir force on a  ball, and was found to give a positive value for $\sigma$.

  Also in the present case we find it  of interest to make a simple numerical check and see what order of magnitude for $c\tau$ results if one inserts~ reasonable physical values for the other quantities present  in the  expression (\ref{A9}).
 Let us choose $\sigma=73~$dyn/cm, the conventional result for an air-water surface, and  choose  $\zeta_p=3\times 10^{16}~$rad/s, a usual value for the plasma frequency. Then, Eq.~(\ref{A9}) yields, when we ignore the sign, the minimum  length to be
 \begin{equation}
 c\tau  \approx 0.4~\rm{\AA}; \label{A110}
 \end{equation}
a number corresponding  to   atomic dimensions. The result is strikingly similar to that obtained in Ref.~\cite{Hoye:2017rst}, although the model considered there was a compact ball instead of a thin shell. One may be tempted to wonder, as we did in Ref.~\cite{Hoye:2017rst}: is there a deeper link between QFT cutoff quantities and common quantities known from hydromechanics?

Keeping $\delta$ constant and omitting $\tau$ in Eq.~(\ref{A7}), we would have obtained, 
for finite coupling,
\begin{equation}
\sigma=-\frac{\zeta_p^2\hbar}{64 \pi c}\frac{1}{a\delta}. 
\end{equation}
This equation is  comparable to Eq.~(\ref{A9}), the arc length $a\delta$ corresponding to the minimum distance $c\tau$.


\subsection{Remarks on isorefractive media}

As is  known, a  surface pressure occurs because of imbalance between the two media separated by a fluid interface: a molecule residing in the interface becomes acted upon by
different forces from neighboring  particles on the inside than from those  on the outside. A noteworthy exception is the case of isorefractive media, where
the product $\varepsilon\mu$ is the same on the two sides. It is illustrative to consider the following simple example: let  two spherical fluid balls 1 and 2 of this sort be touching each externally at one point, identified as the origin of coordinates.  Any disturbance
 in ball 1 at position $\bf r$ will need precisely the same time to reach the origin as a disturbance at the inverted position $-\bf r$ in ball 2. The imbalance becomes in this way eliminated,
 and the effect of surface tension disappears. Mathematically, if one calculates the Casimir  surface pressure on such a ball one finds the  counter term to be  simply zero; there occur no
  divergences in the conventional temporal point-splitting cutoff.  This effect was demonstrated in the detailed calculations in Refs.~\cite{Brevik:1982bpp,Brevik:1982iso,Brevik:1983pt}. The most typical case is when
  \begin{equation}
  \varepsilon \mu=1, \label{A10}
  \end{equation}
  corresponding to a photon velocity in the medium equal to the vacuum value $c$. It is instructive to note the expressions for the interior and exterior
  energies, assuming for simplicity the case where the relative permittivity $\mu_{12}=\mu_1/\mu_2$ is either zero or infinity, in the first order of the UAE:
  \begin{equation}
  E^{(1)}_\mathrm{int}=\frac{\hbar c}{2a}\left[ - \frac{8}{3\pi {\tilde \tau}^2} + \frac{11}{36\pi}+\frac{3}{64} \right],  \label{A11}
  \end{equation}
  \begin{equation}
   E^{(1)}_\mathrm{ext}=\frac{\hbar c}{2a}\left[  \frac{8}{3\pi {\tilde \tau}^2} - \frac{11}{36\pi}+\frac{3}{64} \right]. \label{A12}
   \end{equation}
   Thus $E_\mathrm{int}\rightarrow -\infty$ and $E_\mathrm{ext}\rightarrow +\infty$ when $\tilde{\tau}\rightarrow 0$ but their sum is finite,
   \begin{equation}
   E^{(1)}=E^{(1)}_\mathrm{int}+E^{(1)}_\mathrm{ext}=\frac{3\hbar c}{64a}. \label{A13}
   \end{equation}

 It is also natural here to mention that the condition (\ref{A10}) is a precise analogy to the relativistic  model proposed by Lee for the color medium
 outside a hadron bag~\cite{Lee:1988ppft} with noninteracting gluons playing the role of photons.

  Finally, returning to electrodynamics it is natural to make a comparison with the  case  $\tilde{\lambda}=-\lambda$ encountered in in Eq.~(\ref{isorefractive}). Also in this situation  it turns out to be
   possible,
as noted,  to isolate a unique finite part in the energy. The analogy is not complete, though, since a negative $\lambda$ would imply a negative value of $\zeta_m$ or $\zeta_p$ in Eq.~(\ref{plasma-coupling}).


\bibliography{biblio/delta-sphere}

\end{document}